\documentclass[runningheads,envcountsect,envcountsame]{llncs}
\usepackage{versions}
\excludeversion{LNCS}
\includeversion{ARXIV}
\usepackage[utf8]{inputenc}
\usepackage[T1]{fontenc}

\let\doendproof\endproof
\renewcommand\endproof{~\hfill$\qed$\doendproof}

\usepackage[textsize=scriptsize]{todonotes}%

\usepackage{bm} 
\usepackage{amssymb} 
\usepackage{url}

\usetikzlibrary{arrows.meta,arrows}

\usepackage{mathtools}

\usepackage[frozencache=true,cachedir=minted-cache]{minted}
\setminted[python]{breaklines, framesep=2mm, fontsize=\footnotesize, numbersep=5pt}

\usepackage{cleveref}
\usepackage{xpatch,letltxmacro}

\newcommand{\Nat}{\ensuremath{\mathbb{N}}}
\newcommand{\Rel}{\ensuremath{\mathbb{Z}}}
\newcommand{\tR}{\mathtt{R}}
\newcommand{\ta}{\mathtt{A}}
\newcommand{\tb}{\mathtt{B}}
\newcommand{\tc}{\mathtt{C}}
\newcommand{\calA}{\mathcal A}
\newcommand{\calB}{\mathcal B}
\newcommand{\dwc}{\mathop{\downarrow}\nolimits}
\newcommand{\Cuts}{\operatorname{\textit{Cuts}}}
\newcommand{\alphabet}{\operatorname{\textit{alph}}}

\newcommand{\Id}{\mathit{Id}}
\newcommand{\subword}{\preccurlyeq}
\newcommand{\nsubword}{\not\preccurlyeq}
\newcommand{\subsim}{\lesssim}
\newcommand{\eqby}[1]{\stackrel{\!#1\!}{=}}

\newcommand{\leqby}[1]{\stackrel{\!#1\!}{\leq}}
\newcommand{\eqdef}{\stackrel{\mbox{\begin{scriptsize}def\end{scriptsize}}}{=}}
\newcommand{\equivdef}{\stackrel{\mbox{\begin{scriptsize}def\end{scriptsize}}}{\Leftrightarrow}}
\newcommand{\tuple}[1]{\langle #1 \rangle}
\newcommand{\obracew}[2]{{\overset{#2}{\overbrace{#1}}}}

\title{
       On the piecewise complexity of words and periodic words
\thanks{Work supported by IRL ReLaX. J.\ Veron supported by DIGICOSME ANR-11-LABX-0045.}
}

\author{
        M. Praveen \inst{1,2}
\and    Ph.\ Schnoebelen \inst{3}
\and    J.\ Veron  \inst{3}
\and    I.\ Vialard \inst{3}
}

\institute{
        Chennai Mathematical Institute, Chennai, India
\and    CNRS, ReLaX, IRL 2000, Siruseri, India
\and    Laboratoire M\'ethodes Formelles, Univ.\ Paris-Saclay, France
}

\begin{document}

\maketitle

\begin{abstract}
The piecewise complexity $h(u)$ of a word is the minimal length of subwords needed to exactly characterise $u$. Its piecewise minimality index $\rho(u)$ is the smallest length $k$ such that $u$ is minimal among its order-$k$ class $[u]_k$ in Simon's congruence.

We study these two measures and provide efficient algorithms for computing $h(u)$ and $\rho(u)$. We also provide efficient algorithms for the case where $u$ is a periodic word, of the form $u=v^n$.
\end{abstract}

\section{Introduction}
\label{sec-intro}

For two words $u$ and $v$, we write $u\subword v$ when $u$ is a
\emph{subword}, i.e., a subsequence, of $v$. For example
$\mathtt{SIMON}\subword \mathtt{STIMULATION}$ while $\mathtt{HEBRARD}
\not \subword \mathtt{HAREBRAINED}$. Subwords and subsequences play
a prominent role in many areas of computer science. Our personal
motivations come from descriptive complexity and the possibility of
characterising words and languages via some short witnessing subwords.

Fifty years ago, and with similar motivations, I.~Simon
introduced \emph{piecewise-testable} (PT) languages in his doctoral thesis
(see~\cite{simon72,simon75,sakarovitch83}): a language $L$ is PT if
there is a \emph{finite} set of words $F$
such that the membership of a word $u$ in $L$ depends only on which words from
$F$ are subwords of $u$. PT languages  have since played an
important role in the algebraic and logical theory of first-order
definable languages, see~\cite{pin86,DGK-ijfcs08,klima2011} and the
references therein.  They also constitute an important class of simple
regular languages with applications in learning
theory~\cite{kontorovich2008}, databases~\cite{bojanczyk2012b},
linguistics~\cite{rogers2013}, etc. The concept of PT languages has
been extended to variant notions of ``subwords''~\cite{zetzsche2018},
to trees~\cite{bojanczyk2012b}, infinite
words~\cite{perrin2004,carton2018b}, pictures~\cite{matz98}, or any
combinatorial well-quasi-order~\cite{goubault2016}.

When a PT language $L$ can be characterised via a finite $F$ where all
words have length at most $k$, we say that $L$ is piecewise-testable
\emph{of height $k$}, or $k$-PT.  Equivalently, $L$ is $k$-PT if it is
closed under $\sim_k$, Simon's congruence of order $k$, defined via
$u\sim_k v$ $\equivdef$ $u$ and $v$ have the same subwords of length
at most $k$. The \emph{piecewise complexity} of $L$, denoted $h(L)$ (for
``height''), is the smallest $k$ such that $L$ is $k$-PT. It coincides
with the minimum number of variables needed in any $\calB\Sigma_1$ formula that
defines $L$~\cite{DGK-ijfcs08}.

The piecewise complexity of languages was studied by
Karandikar and Schnoebelen in~\cite{KS-lmcs2019} where it is a central
tool for establishing elementary upper bounds for the complexity of
the $\mathsf{FO}^2$ fragment of the logic of
subwords.

In this paper we focus on the piecewise complexity of
\emph{individual words}. For $u\in A^*$, we write $h(u)$ for
$h(\{u\})$, i.e., the smallest $k$ s.t.\ $[u]_k=\{u\}$, where $[u]_k$
is the equivalence class of $u$ w.r.t.\ $\sim_k$.  We also introduce a
new measure, $\rho(u)$, defined as the smallest $k$ such that $u$ is
minimal in $[u]_k$ (wrt subwords).

We have two main motivations. Firstly
it appeared in~\cite{KS-lmcs2019} that bounding $h(L)$ for a PT
language $L$ relies heavily on knowing $h(u)$ for specific words $u$
in and out of $L$. For example, the piecewise complexity of a
finite language $L$ is exactly $\max_{u\in L}h(u)$~\cite{KS-lmcs2019}, and
the tightness of many upper bounds in~\cite{KS-lmcs2019} relies on
identifying a family of long words with small piecewise
complexity. See also~\cite[Sect.~4]{HS-ipl2019}.
Secondly
the piecewise
complexity of words raises challenging combinatorial or algorithmic
questions. To begin with we do not yet have a practical and efficient
algorithm that computes $h(u)$.

\subsubsection*{Our contribution.}

Along $h(u)$, we introduce a new measure, $\rho(u)$, the \emph{piecewise
minimality index} of $u$, and initiate an investigation of the
combinatorial and algorithmic properties of both measures.  The new
measure $\rho(u)$ is closely related to $h(u)$ but is easier to
compute. Our main results are (1) theoretical results connecting $h$
and $\rho$ and bounding their values in contexts involving
concatenation, (2) efficient algorithms for computing $h(u)$ and
$\rho(u)$, and (3) an analysis of periodicities in the arch
factorization of periodic words that leads to a simple and efficient
algorithm computing $h(u^n)$ and $\rho(u^n)$ for periodic words $u^n$.
Our motivation for computing $h(u^n)$ and $\rho(u^n)$ is that we see
it as preparatory work for computing subword complexity measures on
compressed data, see~\cite{SV-words2023}.

\subsubsection*{Related work.}

In the literature, existing works on $h$ mostly focus on $h(L)$ for
$L$ a PT-language, and provide general bounds (see,
e.g.,~\cite{KS-lmcs2019,HS-ipl2019}). We are not aware of any
practical algorithm computing $h(L)$ for $L$ a PT-language given,
e.g., via a deterministic finite-state automaton $\calA$, and it is
known that deciding whether $h(L(\calA))\leq k$ is
$\textsf{coNP}$-complete~\cite{masopust2015}.

Regarding words, there is a rich literature on algorithms
computing $\delta(u,v)$, the piecewise distance between two words:
see~\cite{simon2003,fleischer2018,barker2020,gawrychowski2021} and the
references therein.  Computing $h(u)$ and $\rho(u)$ amounts to
maximising $\delta(u,v)$ over the set of all words $v$ distinct from $u$
---when computing $h(u)$--- or a subset of these ---for $\rho(u)$---.

\subsubsection*{Outline of the paper.}
After recalling the necessary background in \Cref{sec-basics}, we
define the new measures $h(u)$ and $\rho(u)$ in \Cref{sec-h-rho} and
prove some first elementary properties like monotonicity and
convexity.  In \Cref{sec-algo} we give efficient algorithms computing
$h(u)$ and $\rho(u)$.  In \Cref{sec-arch-and-h-rho} we prove simple
but new connections between Simon's side distance functions $r$,
$\ell$ and H\'ebrard's arch factorization. This motivates the study of
the arch factorization of periodic words and leads to a simple and
efficient algorithm computing $h(u^n)$ and $\rho(u^n)$. 
\begin{LNCS}
For lack of space, most proofs are missing from this extended
abstract: they can be found in the full version of the paper,
available as \texttt{arXiv:2023.01234 [cs.FL]}.
\end{LNCS}



\section{Words, subwords and Simon's congruence}
\label{sec-basics}

We consider finite words $u,v,\ldots$ over a finite alphabet $A$. The
empty word is denoted with $\epsilon$, the mirror (or reverse) of $u$
with $u^\tR$, and we write $|u|$ for the length of $u$. We also write
$|u|_a$ for the number of times the letter $a$ appears in $u$.
For a word $u=a_1a_2\cdots a_L$ of length $L$ we write
$\Cuts(u)=\{0,1,2,\ldots,L\}$ for the set of positions between the
letters of $u$.  For $i\leq j\in\Cuts(u)$, we write $u(i,j)$ for the
factor $a_{i+1}a_{i+2}\cdots a_j$. Note that $u(0,L)=u$,
$u(i_1,i_2)\cdot u(i_2,i_3)=u(i_1,i_3)$ and that $|u(i,j)|=j-i$. We
write $u(i)$ as shorthand for $u(i-1,i)$, i.e., $a_i$, the $i$-th
letter of $u$. With $\alphabet(u)$ we denote the set of letters that
occur in $u$. We often abuse notation and write ``$a\in u$'' instead
of ``$a\in\alphabet(u)$'' to say that a letter $a$ occurs in a word
$u$.

We say that $u=a_1\cdots a_L$ is a \emph{subword} of $v$, written
$u\subword v$, if $v$ can be factored under the form $v=v_0a_1v_1 a_2
\cdots v_{L-1} a_L v_L$ where the $v_i$'s can be any words (and can be
empty).  We write $\dwc u$ for the set of all subwords of $u$: e.g.,
$\dwc\mathtt{ABAA} = \{\epsilon, \mathtt{A}, \mathtt{B}, \mathtt{A A},
\mathtt{A B}, \mathtt{B A}, \mathtt{A A A}, \mathtt{A B A}, \mathtt{B
A A}, \mathtt{A B A A}\}$.

Factors are a special case of subwords: $u$ is a \emph{factor} of $v$
if $v=v'u v''$ for some $v',v''$. Furthermore, when $v=v'u v''$ we say
that $u$ is a \emph{prefix} of $v$ when $v'=\epsilon$, and is a
\emph{suffix} of $v$ when $v''=\epsilon$. 

When $u\neq v$, a word $s$ is a \emph{distinguisher} (or a
\emph{separator}) if $s$ is subword of exactly one word among $u$ and
$v$~\cite{simon72}.

For $k
\in\Nat$ we write $A^{\leq k}$ for the set of words over $A$ that have
length at most $k$, and  for any words $u,v \in A^*$, we let $ u \sim_k v
\equivdef \dwc{u}\cap A^{\leq k}=\dwc{v}\cap A^{\leq k} $.  In other
words, $u\sim_k v$ if $u$ and $v$ have the same subwords of length at
most $k$.  For example $\mathtt{ABAB}\sim_1 \mathtt{AABB}$ (both words
use the same letters) but $\mathtt{ABAB}\not\sim_2 \mathtt{AABB}$
($\mathtt{BA}$ is a subword of $\mathtt{ABAB}$, not of
$\mathtt{AABB}$).
The equivalence $\sim_k$, introduced in~\cite{simon72,simon75}, is called Simon's congruence of order $k$.
Note that $u\sim_0 v$ for any $u,v$, and $u\sim_k u$ for any
$k$. Finally, $u\sim_{k+1} v$ implies $u\sim_k v$ for any $k$, and
there is a refinement hierarchy
${\sim_0}\supseteq{\sim_1}\supseteq{\sim_2}\cdots$ with
$\bigcap_{k\in\Nat}\sim_k=\Id_{A^*}$.
We write $[u]_k$ for the equivalence class of $u\in A^*$ under
$\sim_k$.  Note that each $\sim_k$, for $k=0,1,2,\ldots$, has finite
index~\cite{simon75,sakarovitch83,KKS-ipl2015}.

We further let
$u \subsim_k v \equivdef u\sim_k v \land u\subword v$.
Note that $\subsim_k$ is stronger than $\sim_k$. Both relations are
(pre)congruences: $u\sim_k v$ and $u'\sim_k v'$ imply $uu'\sim_k vv'$,
while $u\subsim_k v$ and $u'\subsim_k v'$ imply $uu'\subsim_k vv'$.

The following properties will be useful:
\begin{lemma}
\label{lem-useful}
For all $u,v,v',w \in A^*$ and $a,b \in A$:
\begin{enumerate}

\item
\label{it-convex}
If  $u \sim_k v$ and  $u\subword w\subword v$ then $u\subsim_k
w\subsim_k v$;

\item
\label{it-carac-richn}
When $k>0$, $u\sim_k u v$ if, and only if, there exists a
factorization $u=u_1 u_2\cdots u_k$ such that $\alphabet(u_1)
\supseteq \alphabet(u_2) \supseteq \cdots \supseteq \alphabet(u_k) \supseteq
\alphabet(v)$;

\item
\label{it-diff-let}
If $u a v\sim_k u b v'$ and $a\neq b$ then
$u bav\sim_k u b v'$ or $u ab v'\sim_k u a v$ (or both);

\item
\label{it-upperbound}
If  $u \sim_k v$ then there exists $w \in
A^*$ such that $u \subsim_k w$ and $v \subsim_k w$;

\item
\label{it-shorter}
If $u\sim_k v$ and $|u|<|v|$ then there exists some $v'$ with
$|v'|=|u|$ and such that $u\sim_k v'\subword v$;

\item
\label{it-pumping}
If $u v \sim_k u a v$ then
$u v \sim_k u a^m v$ for all $m\in\Nat$;

\item
\label{it-sing-inf}
Every equivalence class of $\sim_k$ is a singleton or is infinite.

\end{enumerate}
\end{lemma}
\begin{proof}
(\ref{it-convex}) is by combining
$\dwc{u}\subseteq \dwc{w}\subseteq \dwc{v}$ with the definition of $\sim_k$;
(\ref{it-carac-richn}--\ref{it-upperbound}) are Lemmas 3, 5, and 6 from~\cite{simon75};
(\ref{it-shorter}) is an immediate consequence of Theorem~4 from
\cite[p.~91]{simon72}, showing that all minimal (wrt $\subword$) words in $[u]_k$ have
the same length ---see also
\cite[Theorem~6.2.9]{sakarovitch83} or \cite{fleischer2018}---;
(\ref{it-pumping}) is in the proof of Corollary~2.8 from~\cite{sakarovitch83};
(\ref{it-sing-inf}) follows from (\ref{it-convex}), (\ref{it-upperbound}) and (\ref{it-pumping}).
\end{proof}

The fundamental tools for reasoning about piecewise complexity were
developed in Simon's thesis~\cite{simon72}.
First, there is the concept of ``subword distance''\footnote{In fact
$\delta(u,v)$ is a measure of \emph{similarity} and not of difference,
between $u$ and $v$. The associated distance is actually
$d(u,v)\eqdef 2^{-\delta(u,v)}$~\cite{sakarovitch83}.}
$\delta(u,v)\in\Nat\cup\{\infty\}$,  defined for any $u,v\in A^*$, via
\begin{align}
\label{eq-def-delta}
\delta(u,v)
&\eqdef \max\{k~|~u\sim_k v\}
\\
\label{eq-delta-distinguisher}
&=\begin{cases}
        \infty &\text{if $u=v$,}
        \\
        |s|-1 &\text{if $u\neq v$ and $s$ is a shortest distinguisher.}
\end{cases}
\end{align}
Derived notions are the left and right distances~\cite[p72]{simon72}, defined for any
$u,t\in A^*$, via
\begin{align}
r(u,t)&\eqdef \delta(u,u t) = \max\{k~|~u\sim_k u t\} \:,
\\
\ell(t,u) &\eqdef \delta(t u,u) = \max\{k~|~t u\sim_k u\} \:.
\end{align}

Clearly $r$ and $\ell$ are mirror notions. One usually
proves properties of $r$ only, and (often implicitly) deduce
symmetrical conclusions for $\ell$ by the mirror reasoning.

\begin{lemma}[{\cite[Lemma~6.2.13]{sakarovitch83}}]
\label{lem-uv-uav}
For any words $u,v\in A^*$ and letter $a\in A$
\begin{gather}
\label{eq-uv-uav}
\delta(u v,u a v)=\delta(u,u a)+\delta(a v,v)=r(u,a)+\ell(a,v)
\:.
\end{gather}
\end{lemma}



\section{The piecewise complexity of words}
\label{sec-h-rho}

In this section we define the complexity measures $h(u)$ and
$\rho(u)$, give characterisations in terms of the side
distance functions  $r$ and $\ell$, compare the two measures and establish some first results on the measures of concatenations.

\subsection{Defining words via their subwords}
\label{ssec-h-def}

The piecewise complexity of PT languages was defined
in~\cite{KS-csl2016,KS-lmcs2019}. Formally, for a language $L$ over
$A$, $h(L)$ is the smallest index $k$ such that $L$ is
$\sim_k$-saturated, i.e., closed under $\sim_k$.
For a word $u\in A^*$, this becomes $h(u) \eqdef \min\{n~|~ \forall
v:u\sim_n v\implies u=v\}$: we call it the \emph{piecewise complexity} of
$u$.

\begin{proposition} For any $u\in A^*$,
\label{prop-hu-delta-u-u1au2}
\begin{align}
\label{eq-hu-delta-u-u1au2}
h(u)
&= \max_{\substack{u=u_1u_2 \\ a\in A}} \delta(u,u_1a u_2)+1
\\
\label{eq-hu-delta-u1-u1a-u2-au2}
\tag{H}
&= \max_{\substack{u=u_1u_2 \\ a\in A}}r(u_1,a)+\ell(a,u_2)+1
\:.
\end{align}
\end{proposition}
\begin{ARXIV}
\begin{proof}
We only prove \eqref{eq-hu-delta-u-u1au2} since
\eqref{eq-hu-delta-u1-u1a-u2-au2}  is just a
rewording based on \Cref{eq-uv-uav}.  \\
$(\geq)$: By definition, $h(u)>\delta(u,v)$ for any $v\neq u$, and in
particular for any $v$ of the form $v=u_1 a u_2$.  \\
$(\leq)$: Let $k=h(u)-1$. By definition of $h$, there exists some
$v\neq u$ with $u\sim_k v$. By
\Cref{lem-useful}~(\ref{it-upperbound}), we can further assume
$u\subsim_k v$, and by \Cref{lem-useful}~(\ref{it-convex}), we can
even assume that $|v|=|u|+1$, i.e., $v = u_1 a u_2$ for some $a\in A$
and some factorization $u=u_1u_2$.  Now $\delta(u,v)=\delta(u,u_1 a
u_2)\geq k$ since $u\sim_k v$.  Thus $h(u)=k+1\leq \delta(u,u_1 a
u_2)+1$ for this particular choice of $u_1,u_2$ and $a$.
\end{proof}
\end{ARXIV}

\subsection{Reduced words and the minimality index}
\label{ssec-rho-def}

\begin{definition}[{\cite[p.~70]{simon72}}]
Let $m>0$, a word $u \in A^{*}$ is \emph{$m$-reduced} if
$u\not\sim_m u'$ for all strict subwords $u'$ of $u$.
\end{definition}
In other words, $u$ is $m$-reduced when it is a minimal word in
$[u]_{m}$.
This leads to a new piecewise-based measure for words, that we call
the \emph{minimality index}:
\begin{equation}
\rho(u) \eqdef \min \{ m ~|~ u \text{ is $m$-reduced} \}
\:.
\end{equation}

\begin{lemma}[{\cite[p.~72]{simon72}}]
\label{lem-reduced}
A non-empty word $u$ is $m$-reduced iff $r(u_1,a)+\ell(a,u_2) < m$ for all
factorizations $u=u_1 a u_2$ with $a\in A$ and $u_1,u_2\in A^*$.
\end{lemma}
\begin{ARXIV}
\begin{proof}
Assume, by way of contradiction, that $r(u_1,a)+\ell(a,u_2)\geq m$ for
some factorization $u=u_1 a u_2$.
\Cref{lem-uv-uav} then gives $\delta(u,u_1u_2)\geq
m$, i.e.,  $u\sim_m u_1 u_2$. Finally $u$ is not
minimal in $[u]_m$.
\end{proof}
\end{ARXIV}
This has an immediate corollary:
\begin{proposition} For any non-empty word $u\in A^*$
\label{prop-rho-charac}
\begin{align}
\label{eq-m-via-r&l}
\tag{P}
\rho(u) = \max_{\substack{u=v_1av_2\\ a\in A}} r(v_1,a) + \ell(a,v_2) + 1
\:.
\end{align}
\end{proposition}
Note the difference between \Cref{eq-hu-delta-u1-u1a-u2-au2,eq-m-via-r&l}: $h(u)$ can be computed by looking at all ways 
one would insert a letter $a$ inside $u$ while for $\rho(u)$ one is looking at
all ways one could remove some letter from $u$.

\subsection{Fundamental properties of side distances}

The characterisations given in
\Cref{prop-hu-delta-u-u1au2,prop-rho-charac} suggest that
computing $h(u)$ and $\rho(u)$ reduces to computing the $r$ and $\ell$
side distance functions on prefixes and suffixes of $u$. This will be
confirmed in \Cref{sec-algo}.

For this reason we now prove some useful combinatorial results on $r$
and $\ell$. They will be essential for proving more general properties of $h$
and $\rho$ in the rest of this section, and in the analysis of algorithms
in later sections.
\begin{lemma}
For any word $u\in A^*$ and letters $a,b\in A$:
\begin{gather}
\label{eq-ruba-leq-1+rub}
r(u a,b) \leq 1+r(u,a)\:.
\end{gather}
\end{lemma}
\begin{ARXIV}
\begin{proof}
By definition $r(u,a)$ is $|s|-1$ for $s$ a shortest distinguisher of
$u$ and $ua$.  Necessarily $s\subword ua$ while $s\not\subword u$. So
$sb\subword uab$ and $sb\not\subword ua$, i.e., $sb$ distinguishes
between $ua$ and $uab$, proving $r(ua,b)\leq |sb|-1=|s|=1+r(u,a)$.
\end{proof}
\end{ARXIV}
The following useful lemma provides a recursive way of computing
$r(u,t)$.
\begin{lemma}[{\cite[p.~71--72]{simon72}}]
For any $u,t\in A^*$ and $a\in A$:
\label{lem-sld-comp}
\begin{align}
\label{eq-sld-t}
\tag{R1}
r(u,t) &= \min \bigl\{r(u, a) ~|~ a\in\alphabet(t)\bigr\} \:,
\\
\label{eq-sld-0}
\tag{R2}
r(u,a) &= 0 \text{ if $a$ does not occur in $u$} \:,
\\
\label{eq-sld-u'}
\tag{R3}
r(u,a) &= 1 + r(u',a u'') \text{ if $u=u' a u''$ with
$a\not\in\alphabet(u'')$} \:.
\end{align}
\end{lemma}
\begin{ARXIV}
\begin{proof}
\eqref{eq-sld-t}: Since $r(u,\epsilon)=\delta(u,u)=+\infty=\min\emptyset$,
the statement holds when $t=\epsilon$. So we may assume $t\neq
\epsilon$ and $r(u,t)\in\Nat$.

Pick $a$ occurring in $t$ and let $k=1+r(u,a)$. By
\Cref{eq-delta-distinguisher}, there exists $s$ of length $k$ such
that $s\nsubword u$ and $s\subword u a$. Hence $s\subword u t$, and $s$
is a distinguisher for $u$ and $u t$. With Eq.~\eqref{eq-delta-distinguisher} we deduce $r(u,
t)\leq |s|-1=r(u,a)$. Since this
holds for any $a$ in $t$, we conclude
$r(u,t)\leq\min_{a\in\alphabet(t)}r(u,a)$.

For the ``$\geq$'' direction, let $k=1+r(u, t)$ and pick a
distinguisher $s$ of length $k$ such that $s\nsubword u$ and
$s\subword u t$.  Write $s=s_1a s_2$ with $s_1$ the longest prefix of
$s$ that is a subword of $u$, $a$ the first letter after $s_1$, and
$s_2$ the rest of $s$. Now $a s_2\subword t$ so that
$a\in\alphabet(t)$. Since $s_1a\nsubword u$, we deduce $r(u, a)\leq
|s_1a|-1\leq |s|=1+r(u, t)$. We have then found some $a\in\alphabet(t)$
with $r(u,t)\geq r(u,a)$.  \\

\eqref{eq-sld-0}:
If $a$ does not occur in $u$, it is a distinguisher with $u a$
hence $r(u,a)=0$ by \Cref{eq-delta-distinguisher}.
\\

\eqref{eq-sld-u'}:
Assume $|u|_a>0$ and write $u=u' a u''$ with $|u''|_a=0$.

Let $k=1+r(u',a u'')$. By \Cref{eq-delta-distinguisher} there exists
a distinguisher $s$ of length $k$ with $s\nsubword u'$ and $s\subword
u' a u''$, further entailing $s a\nsubword u' a u''=u$ and $s a\subword
u' a u'' a=u a$, i.e., $s a$ is a distinguisher for $u$ and $u a$.  We
deduce $r(u,a)\leq |s a|-1=k=1+r(u',a u'')$, proving the ``$\leq$''
direction of \eqref{eq-sld-u'}.

For the other direction, let $k=r(u,a)$. Then there is a distinguisher
$s$ of length $k+1$ with $s\nsubword u$ and $s\subword
u a$. Necessarily $s$ is some $ta$, with $t\subword u$. From
$ta\nsubword u$, we deduce $t\nsubword u'$ and then $t$ distinguishes
between $u'$ and $u=u' a u''$. Then $r(u',a u'')\leq |t|-1\leq k-1$,
i.e., $1+r(u',a u'')\leq k=r(u,a)$.
\end{proof}
\end{ARXIV}

\begin{corollary}
\label{coro-monot-subalphabet}
\label{coro-dr-a-notin-v}
For any $u,v,t,t'\in A^*$ and $a\in A$
\begin{gather}
\label{eq-monot-subalphabet}
\alphabet(t)\subseteq \alphabet(t')
\implies
r(u,t)\geq r(u,t') \text{ and }
\ell(t,u)\geq \ell(t',u)
\:,
\\
\label{eq-dr-a-notin-v}
a\not\in\alphabet(v)
\implies r(u v,a) \leq r(u,a) \text{ and }
\ell(a,v u) \leq \ell(a,u)
\:.
\end{gather}
\end{corollary}
\begin{ARXIV}
\begin{proof}
We only prove the claims on $r$ since the claim on $\ell$ can be
derived by mirroring.
The implication in~\eqref{eq-monot-subalphabet} is direct
from~\eqref{eq-sld-t}.
For the proof of \eqref{eq-dr-a-notin-v}, the assumption is that $a$
does not occur in $v$ and we consider two cases:

(i)
if  $a$ does not occur in $u$, we have
$r(u v,a) \eqby{\eqref{eq-sld-0}} 0
\eqby{\eqref{eq-sld-0}}r(u,a)$.

(ii) if $a$ occurs in $u$, we write $u=u' a u''$ with $a$ not
occurring in $u''$, so that $r(u v,a)
\eqby{\eqref{eq-sld-u'}}
1+r(u',a u''v)
\leqby{\eqref{eq-monot-subalphabet}}
1+r(u',a u'')
\eqby{\eqref{eq-sld-u'}}
r(u,a)$.
\end{proof}
\end{ARXIV}

\begin{lemma}[Monotonicity of $r$ and $\ell$]
\label{lem-sld-mono}
For all $u,v,t\in A^*$
\begin{xalignat}{2}
\label{eq-sld-mono}
r(v,t)&\leq r(u v,t)
\:,
&
\ell(t,u)&\leq \ell(t,u v)
\:.
\end{xalignat}
\end{lemma}
\begin{ARXIV}
{\renewcommand\endproof{\doendproof}
\begin{proof}
We prove the left-side inequality by induction on $|v|$ and then on
$|t|$, the right-side inequality being derived by mirroring. We
consider several cases:

(i) if $t=\epsilon$ then $r(v,t) = \infty = r(u v,t)$.

(ii) if $t$ is a letter not occurring in $v$, then
$r(v,t)\eqby{\eqref{eq-sld-0}}0\leq r(u v,t)$.

(iii) if $t=a$ is a letter occurring in $v$, we write $v$ under the
form $v=v' av''$ with $a$ not occurring in $v''$ and derive 
\begin{gather*}
r(v,t)
\eqby{\eqref{eq-sld-u'}} 1+r(v',av'') \leqby{\text{i.h.}}
1+r(u v',av'') \eqby{\eqref{eq-sld-u'}} r(u v,t)
\:.
\end{gather*}

(iv) if $|t|>1$ then
\begin{gather}
\tag*{\qed}
r(v,t) \eqby{\eqref{eq-sld-t}}
\min_{a\in t}r(v,a) \leqby{\text{i.h.}}
\min_{a\in t}r(u v,a) \eqby{\eqref{eq-sld-t}}
r(u v,t)
\:.
\end{gather}
\end{proof}}
\end{ARXIV}

Observe that $r(u,a)$ can be strictly larger than $r(u v,a)$, e.g.,
$r(a a,a) = 2 > r(a a b,a) = 1$. However, inserting a letter in $u$
cannot increase $r$ or $\ell$ by more than one:
\begin{lemma}
\label{lem-mono-insert-1a}
For any $u,v\in A^*$ and $a,b\in A$
\begin{xalignat}{2}
\label{eq-mono-insert-1a}
r(u a v,b)&\leq 1+r(u v,b) \:,
&
\ell(b,u a v)&\leq 1+\ell(b,u v)  \:.
\end{xalignat}
\end{lemma}
\begin{ARXIV}
\begin{proof}
We prove the first inequality by induction on $|v|$, the second
inequality being derived by mirroring. We consider several cases:
\begin{enumerate}
\item
if $b$ does not occur in $u a v$ then
$r(u a v,b)=0=r(u v,b)$.

\item
if $b$ occurs in $v$, we write $v=v' b v''$ with $b$ not occurring in
$v''$ and derive
\begin{align*}
r(u a v,b)
&\eqby{\eqref{eq-sld-u'}}
1+r(u a v',b v'')
\eqby{\eqref{eq-sld-t}}
1+\min_{c\in b v''}r(u a v',c)\\
&\leqby{\text{i.h.}}
2+\min_{c\in b v''}r(u v',c)
\eqby{\eqref{eq-sld-t}}
1+r(u v' b v'',b)=
1+r(u v,b)
\:.
\end{align*}

\item
if $b$ occurs in $u$ but not in $av$, we write $u=u' bu''$ with $b$ not
occurring in $u''$ and derive
$r(u a v,b)
\eqby{\eqref{eq-sld-u'}}
1+r(u',bu'' av)
\leqby{\eqref{eq-monot-subalphabet}}
1+r(u',bu''v)
\eqby{\eqref{eq-sld-u'}}
r(u v,b)$.

\item
if $b=a$ does not occur in $v$ nor in $u$, we have
\begin{gather*}
r(u a v,b)
\eqby{\eqref{eq-sld-u'}}
1+r(u,av)
\leqby{\eqref{eq-monot-subalphabet}}
1+r(u,a)
\eqby{\eqref{eq-sld-0}} 1
\:.
\end{gather*}

\item
if $b=a$ does not occur in $v$ but occurs in $u$, we write $u=u' a u''$
with $a$ not occurring in $u''$.  The left-hand term of the inequality
can be rewritten
\begin{align*}
r(u a v,b)
\eqby{\eqref{eq-sld-u'}}
1+r(u,av)
\eqby{\eqref{eq-sld-t}}
&\min[1+r(u,a),1+r(u,v)]
\\
\eqby{\eqref{eq-sld-u'}}
&\min[\obracew{2+r(u_1,a u_2)}{d_1},\obracew{1+r(u,v)}{d_2}]
\:.
\shortintertext{For the right-hand term, we have}
1+r(u v,b)
\eqby{\eqref{eq-sld-u'}}
2+r(u_1,a u_2v)
\eqby{\eqref{eq-sld-t}}
&\min[\obracew{2+r(u_1,a u_2)}{d_1},\obracew{2+r(u_1,v)}{d_3}]
\:.
\end{align*}
If now $d_1\leq d_3$, we are done. So assume $d_3<d_1$. By
Eq.~\eqref{eq-sld-t}, this implies that $r(u_1,v)=r(u_1,c)$ for
some letter $c$ in $v$ that does not occur in $a u_2$. We derive $d_2
\leqby{\eqref{eq-monot-subalphabet}} 1+r(u,c)
\leqby{\eqref{eq-dr-a-notin-v}} 1+r(u_1,c)=d_3-1$ which proves the
inequality.
\end{enumerate}
All possibilities for $u,v,a,b$ have been accounted for.
\end{proof}
\end{ARXIV}

\subsection{Relating $h$ and $\rho$}

\begin{theorem}
\label{thm-h-m}
$h(u) \geq 1+ \rho(u)$ for any word $u$.
\end{theorem}
\begin{ARXIV}
\begin{proof}
For the empty word, one has $h(\epsilon)=1$ and $\rho(\epsilon)=0$.
We now assume that $u$ is non-empty:
by \Cref{prop-rho-charac}, there is a factorization $u=u_1 a u_2$ with
$a\in A$ such that
\begin{align*}
\rho(u) &= 1 + r(u_1,a) + \ell(a,u_2)
\eqby{\eqref{eq-sld-u'}} r(u_1a,a)+\ell(a,u_2)
\leqby{\eqref{eq-hu-delta-u1-u1a-u2-au2}}
h(u)-1
\:.
\end{align*}
and this concludes the proof.
\end{proof}
\end{ARXIV}

The above inequality is an equality in the special case of binary words.
\begin{theorem}
\label{thm-h-m-2letter}
Assume  $|A|= 2$.  Then  $h(u) = \rho(u) + 1$ for any $u\in A^*$.
\end{theorem}
\begin{ARXIV}
\begin{proof}
In view of \Cref{thm-h-m}, there only remains to prove $h(u)\leq \rho(u)+1$.
Let us fix $A=\{a,b\}$ w.l.o.g.\ and assume $u\in A^*$.
Using \eqref{eq-hu-delta-u1-u1a-u2-au2} we know that
$h(u)=1+r(u_1,x)+\ell(x,u_2)$ for some letter $x\in A$ and
factorization
$u=u_1u_2$. W.l.o.g.\ we can assume $x=a$ and consider  several
possibilities
for $u_1$ and $u_2$.
\begin{description}

\item[Case 1:] $u_1$ ends with $a$, i.e., $u_1 = u'_1 a$. Then
\begin{equation}
\begin{aligned}
\notag
 h(u) &= 1+r(u'_1 a,a) +\ell(a,u_2) \eqby{\eqref{eq-sld-u'}} 2+r(u'_1,a) + \ell(a,u_2) \\
	 &\leqby{\eqref{eq-m-via-r&l}} 1+\rho(u'_1a u_2)=1+\rho(u) \:.
\end{aligned}
\end{equation}
\item[Case 2:] $u_1$ ends with $b$ and $u_2$ starts with $b$, i.e.,
  $u_1 = u'_1 b$, $u_2 = bu'_2$, and $u= u'_1 bb u'_2$. Then
\begin{equation}
\begin{aligned}
\notag
 h(u) &= 1+r(u_1,a) +\ell(a,u_2) = 1+r(u'_1 b,a) +\ell(a,bu'_2)
\\
      &\leqby{\eqref{eq-ruba-leq-1+rub}} 1 + (1+r(u'_1,b)) + (1+\ell(b,u'_2))
      = 3 + r(u'_1,b) + \ell(b,u'_2)
\\
      &\eqby{\eqref{eq-sld-u'}} 2+r(u'_1b,b) + \ell(b,u'_2) \leqby{\eqref{eq-m-via-r&l}} 1+\rho(u'_1b b u'_2)=1+\rho(u) \:.
\end{aligned}
\end{equation}
\item[Case 3:] $u_1$ ends with $b$ and $u_2$ is	 empty, i.e.,
  $u_1 = u'_1 b$ and $u=u_1$. Then
\begin{equation}
\begin{aligned}
\notag
 h(u) &= 1+r(u_1,a) +\ell(a,u_2) = 1+r(u'_1 b,a) +\ell(a,\epsilon)
\\
	&\eqby{\eqref{eq-sld-0}} 1+r(u'_1 b,a) +\ell(b,\epsilon) \leqby{\eqref{eq-ruba-leq-1+rub}} 1 + 1+r(u'_1,b) + \ell(b,\epsilon)
\\
      &\leqby{\eqref{eq-m-via-r&l}} 1+\rho(u'_1 b u'_2)=1+\rho(u) \:.
\end{aligned}
\end{equation}
\item[Remaining Cases:] If $u_2$ starts with $a$, we use symmetry and
  reason as in Case 1.	If $u_2$ starts with $b$ and $u_1$ is empty,
  we reason as in Case 2.  If both $u_1$ and $u_2$ are empty, we have
  $u=\epsilon$, $h(\epsilon)=1$ and $\rho(\epsilon)=0$.
\end{description}
All cases have been dealt with and the proof is complete.
\end{proof}
\end{ARXIV}

\begin{remark}
\Cref{thm-h-m-2letter} cannot be generalised to words using $3$ or more
different letters.  For example, with $A=\{\ta,\tb,\tc\}$ and $u=\mathtt{CAACBABA}$,
one has $h(u)=5$ and $\rho(u)=3$.  Larger gaps are possible:
$u=\mathtt{CBCBCBCBBCABBABABAAA}$ has $h(u)=10$ and $\rho(u)=6$.
\qed
\end{remark}

\subsection{Subword complexity and concatenation}
\label{ssec-subw-compl-concat}

While the subwords of $u v$ are obtained by concatenating the subwords
of $u$ and the subwords of $v$, there is no simple relation between
$h(u v)$ or $\rho(u v)$ on one hand, and $h(u)$, $h(v)$, $\rho(u)$ and
$\rho(v)$ on the other hand.

However, we can prove that $h$ and $\rho$ are monotonic and convex
wrt concatenation.

We start with convexity.
\begin{theorem}[Convexity]
\label{thm-add-h-rho}
For all $u,v\in A^*$
\begin{xalignat}{2}
\label{eq-add-rho-h}
\rho(u v)&\leq \rho(u)+\rho(v)
\:,
&
h(u v)&\leq \max \bigl\{ h(u)+\rho(v), \: \rho(u)+h(v)\bigr\}
\:.
\end{xalignat}
\end{theorem}
Note that the second inequality entails
$h(u v)\leq h(u)+h(v) -1$ and is in fact stronger.

\begin{ARXIV}
The proof of \Cref{thm-add-h-rho} relies on the following two lemmas:
\begin{lemma}
For any $u\in A^*$, for any $a\in A$
\begin{equation}
\label{eq-r-leq-rho}
r(u,a)\leq\rho(u)\:.
\end{equation}
\end{lemma}
{\renewcommand\endproof{\doendproof}
\begin{proof}
If $a$ does not appear in $u$, then $r(u,a)\eqby{\eqref{eq-sld-0}} 0$.
On the other hand,
if $u = u' a u''$ with $a \not\in \alphabet(u'' )$, then
\begin{align}
\notag
r(u,a) & \eqby{\eqref{eq-sld-u'}} r(u',a u'') + 1 \eqby{\eqref{eq-sld-t}}
\min \bigl\{r(u', b) ~|~ b\in\alphabet(a u'')\bigr\}+1
\\
\tag*{\qed}
       & \leq r(u',a)+1 \leq r(u',a)+\ell(a,u'') +1 \leqby{\eqref{eq-m-via-r&l}} \rho(u)
\:.
\end{align}
\end{proof}}

\begin{lemma}
For any $u,v,t\in A^*$
\begin{xalignat}{2}
\label{eq-sld-extend3}
r(u v,t)&\leq \rho(u) + r(v,t)
\:,
&
\ell(t,u v)&\leq \rho(v) + \ell(t,u)
\:.
\end{xalignat}
\end{lemma}
{\renewcommand\endproof{\doendproof}
\begin{proof}
We prove the first inequality by induction on $|v|$ and then on $|t|$,
the second inequality being derived by mirroring. We consider several
cases:

(i) if $t=\epsilon$ then $r(u v,t) = \infty = r(v,t)$.

(ii) if $t$ is a letter not occurring in $v$, then $r(u v,t)
\leqby{\eqref{eq-dr-a-notin-v}} r(u,t)
\leqby{\eqref{eq-r-leq-rho}} \rho(u)$.

(iii) if $t=a$ is a letter occurring in $v$, we write $v$
under the form $v =v' a v''$ with $a\not\in v''$ and
derive 
\begin{gather*}
r(u v,a) \eqby{\eqref{eq-sld-u'}} r(u v',av'')+1
\leqby{\text{i.h.}}  r(v',av'') + 1 + \rho(u) \eqby{\eqref{eq-sld-u'}}
r(v,a)+\rho(u)\:.
\end{gather*}

(iv) if $|t|>1$ then we have
\begin{gather}
\tag*{\qed}
r(u v,t) \eqby{\eqref{eq-sld-t}}
\min_{a\in t} r(u v,a) \leqby{\text{i.h.}}
\min_{a\in t}r(v,a)+\rho(u) \eqby{\eqref{eq-sld-t}}
r(v,t)+\rho(u)
\:.
\end{gather}
\end{proof}
}
\begin{proof}[of \Cref{thm-add-h-rho}]
For $\rho(u v)$ we assume that $u v\neq\epsilon$,
otherwise the claim is trivial.
Now by \Cref{prop-rho-charac}, $\rho(u v)$ is
$r(w_1,a)+\ell(a,w_2)+1$ for some letter $a$ and some
factorization $u v=w_1 a w_2$ of $u v$.
Let us assume w.l.o.g.\ (the other case
is symmetrical) that $w_1=u_1$ is a prefix of $u=u_1 a u_2$ so that
$w_2=u_2v$.
Now 
\begin{gather*}
\rho(u)= r(u_1,a)+\ell(a,u_2v)+1
\leqby{\eqref{eq-sld-extend3}} r(u_1,a)+\ell(a,u_2)+1 + \rho(v)
\leqby{\eqref{eq-m-via-r&l}} \rho(u) + \rho(v)
\:.
\end{gather*}
Similarly for $h$: By \Cref{prop-hu-delta-u-u1au2}, $h(u v)$ is
$r(w_1,a)+\ell(a,w_2)+1$ for some letter $a$ and some factorization $u
v=w_1w_2$ of $u v$. If $w_1=u_1$ is a prefix of $u=u_1u_2$ so that
$w_2=u_2v$, then 
\begin{gather*}
h(u v)= r(u_1,a) + \ell(a,u_2v) + 1
\leqby{\eqref{eq-sld-extend3}} r(u_1,a) + \ell(a,u_2) + \rho(v)+1
\leqby{\eqref{eq-hu-delta-u1-u1a-u2-au2}} h(u) + \rho(v) \:.
\end{gather*}
On the
other hand, if $w_2$ is a suffix of $v$, through the same reasoning we
get $h(u v)\leq \rho(u)+h(v)$.
\end{proof}
\end{ARXIV}






\begin{theorem}[Monotonicity]
\label{thm-mono-h}
For all $u,v\in A^*$
\begin{xalignat}{2}
\label{eq-mono-h}
h(u)&\leq h(u v)
\:,
&
h(v)&\leq h(u v)
\:,
\\
\label{eq-mono-rho}
\rho(u)&\leq \rho(u v)
\:,
&
\rho(v)&\leq \rho(u v)
\:.
\end{xalignat}
\end{theorem}
\begin{ARXIV}
\begin{proof}
By \Cref{prop-hu-delta-u-u1au2}, $h(u)$ is $\delta(u_1u_2,u_1a u_2)+1$
for some letter $a\in A$ and some factorization $u=u_1 u_2$ of $u$. We
derive 
\begin{gather*}
h(u)\eqby{\eqref{eq-hu-delta-u1-u1a-u2-au2}}
r(u_1,a)+\ell(a,u_2)+1 \leqby{\eqref{eq-sld-mono}}
r(u_1,a)+\ell(a,u_2v)+1 \leqby{\eqref{eq-hu-delta-u1-u1a-u2-au2}}
h(u v)
\:.
\end{gather*}
The second inequality in Eq.~\eqref{eq-mono-h} is derived by
mirroring, and Eq.~\eqref{eq-mono-rho} is proved in similar way.
\end{proof}
\end{ARXIV}

\section{Computing $h(u)$ and $\rho(u)$}
\label{sec-algo}

Thanks to \Cref{prop-hu-delta-u-u1au2,prop-rho-charac}, computing $h$
and $\rho$ reduces to computing $r$ and $\ell$.
For $r$ and $\ell$ we may rephrase
\Cref{lem-sld-comp} in the following recursive form:
\begin{equation}
\label{eq-rec-r-1}
r(u,a)=
\begin{cases}
  \displaystyle 0
	& \text{if $a\not\in u$},
  \\
  \displaystyle 1+\min_{b\in a u_2} r(u_1,b)
	& \text{if $u=u_1a u_2$ and $a\not\in u_2$}.
\end{cases}
\end{equation}
with a mirror formula for $\ell$.

We now derive a reformulation that leads to more efficient
algorithms.
\begin{lemma}
\label{lem-sld-dynpro}
For any word $u\in A^*$ and letters $a,b\in A$
\begin{equation}
\label{eq-sld-dynpro}
r(u b,a) =
\begin{cases}
        0 & \text{if $a\not\in u b$,}
\\
        1 + r(u,a) & \text{if $a=b$,}
\\
        \min\left\{\!\begin{array}{c}
               1+r(u_1,b)\\
                r(u,a)
      \end{array}\!\right\}\!\!\!
     &\text{if $a\neq b$ and $u=u_1 a u_2$ with $a\not\in u_2$.}
\end{cases}
\end{equation}
\end{lemma}
\begin{ARXIV}
{\renewcommand\endproof{\doendproof}
\begin{proof}
When $a$ does not occur in $u b$ or when $a=b$, Eq.~\eqref{eq-sld-0} or
Eq.~\eqref{eq-sld-u'} directly gives $r(u b,a)=0$ or
$r(u b,a)=1+r(u,a)$.
So assume $a\neq b$ and write $u=u_1 a u_2$ with $a$ not occurring in
$u_2$.	Using \Cref{eq-sld-u',eq-sld-t} give
\begin{align}
\notag
r(u b,a) & = r(u_1 a u_2 b,a)
\\
\notag
&\eqby{\eqref{eq-sld-u'}} 1+r(u_1,a u_2 b)
\eqby{\eqref{eq-sld-t}} 1+\min_{c\in a u_2 b}r(u_1,c)
\\
\notag
&=
1+\min\bigl[r(u_1,b),\min_{c\in a u_2}r(u_1,c)\bigr]
\\
\tag*{\qed}
&\eqby{\eqref{eq-sld-t}}
\min [1+r(u_1,b),1+r(u_1,a u_2)]
\eqby{\eqref{eq-sld-u'}} \min [1+r(u_1,b),r(u,a)]
\:.
\end{align}
\end{proof}
}
\end{ARXIV}

The recursion in \Cref{eq-rec-r-1} involves the prefixes
of $u$. We define the \emph{$r$-table of $u$} as the rectangular
matrix containing all the $r(u(0,i),a)$ for $i=0,1,\ldots,|u|$ and
$a\in A$. In practice we write just $r(i,a)$ for $r(u(0,i),a)$.
\begin{example}
\label{ex-w1}
Let $u=\mathtt{ABBACCBCCABAABC}$ over
$A=\{\ta,\tb,\tc\}$. The $r$-table of $u$ is
{\excludeversion{WITHARCHES}
\begin{center}
\scalebox{1.0}{
\begin{tikzpicture}[auto,x=5mm,y=3mm,anchor=mid,baseline,node distance=1.3em]
{
\def\Wc{0.92}
\def\ixh{2.7}
\def\alphah{2.9}
\def\wlh{1.5}
\def\wlet{\ta, \tb, \tb, \ta, \tc, \tc, \tb, \tc, \tc,
\ta, \tb, \ta, \ta, \tb, \tc}
\def\riA{0, 1, 1, 1, 2, 1, 1, 1, 1, 1, 2, 2, 3, 4, 4, 3}
\def\riB{0, 0, 1, 2, 2, 1, 1, 2, 2, 2, 2, 3, 3, 3, 4, 3}
\def\riC{0, 0, 0, 0, 0, 1, 2, 2, 3, 4, 2, 2, 2, 2, 2, 3}

\node[left] at (-1.0,\ixh) {\tiny $i$};
\foreach \i in {0,...,15} \node at ({\i*\Wc},\ixh) {\tiny \i};

\node[left] at (-1.0,\wlh) {$w$};
\foreach \val [count=\i] in \wlet \node at ({(\i-0.5)*\Wc},\wlh) {$\val$};

\node[left] at (-1.0,0) {$r(i,\ta)$};
\foreach \val [count=\i] in \riA \node at ({(\i-1)*\Wc},0) {\val};

\node[left] at (-1.0,-1.3) {$r(i,\tb)$};
\foreach \val [count=\i] in \riB \node at ({(\i-1)*\Wc},-1.3) {\val};

\node[left] at (-1.0,-2.6) {$r(i,\tc)$};
\foreach \val [count=\i] in \riC \node at ({(\i-1)*\Wc},-2.6) {\val};

{\tikzstyle{every path}=[draw,thick,-]
\path (-0.8,\ixh) -- (-0.8,-2.9);
\path (-3.2,0.85) -- (15.5*\Wc-0.15,0.85);
}

\begin{WITHARCHES}
\foreach \i in {0, 5, 10, 15}
\fill[rounded corners,very nearly transparent] (-0.4+\i*\Wc,0.8) rectangle (0.4+\i*\Wc,-3.4);
\end{WITHARCHES}

\begin{WITHARCHES}
{\tikzstyle{every path}=[thin,-Latex,shorten <=1mm]
\def\xtr{0.3}
\path (0*\Wc,\alphah) edge[bend left=30] node [above] {$\alpha$} (5*\Wc,\alphah);
\path (5*\Wc,\alphah) edge[bend left=30] (10*\Wc,\alphah);
\path (10*\Wc,\alphah) edge[bend left=30] (15*\Wc,\alphah);
}
\end{WITHARCHES}

}
\end{tikzpicture}

}
\end{center}}
As an exercise in reading \Cref{eq-sld-dynpro}, let us check that the
values in this $r$-table are correct.  First $r(0,a)=0$ for all
letters $a\in A$ since $a$ does not occur in $u(0,0)=\epsilon$ which
is empty.  Since $u(0,1)=\ta$ does not contain $\tb$, we further have
$r(1,\tb)=0$.  And since $u(0,4)=\mathtt{ABBA}$ does not contain
$\tc$, we have $r(i,\tc)=0$ for all $i=0,\ldots,4$.

Let us now check, e.g., $r(6,a)$ for all $a\in A$.  Since
$u(0,6)=\mathtt{ABBACC}$ ends with $b=\tc$, the second case in
Eq.~\eqref{eq-sld-dynpro} gives $r(6,\tc)=1+r(5,\tc)=1+1=2$.
For $a=\ta$, we find that the last occurrence of $\ta$ in $u(0,6)$ is
at position $4$. So $r(6,\ta)$ is the minimum of $r(5,\ta)$ and
$1+r(3,\tc)$, which gives 1. For $a=\tb$, and since $\tb$ last occurs
at position $3$ in $u(0,6)$, $r(6,\tb)$ is obtained as $\min
\bigl(r(5,\tb),1+r(2,\tc)\bigr)=\min (1,1+0)=1$.
\qed
\end{example}

It is now clear that \Cref{eq-sld-dynpro}
directly leads to a $O(|A|\cdot |u|)$ algorithm
for computing $r$-tables. The following code builds the table
from left to right. While progressing through $i=0,1,2,\ldots$, it
maintains a table $\texttt{locc}$ storing, for each $a\in A$, the
position of its last occurrence so far.
\LetLtxMacro{\cminted}{\minted}
\let\endcminted\endminted
\xpretocmd{\cminted}{\RecustomVerbatimEnvironment{Verbatim}{BVerbatim}{}}{}{}
\begin{center}
\begin{cminted}{python}
'''Algorithm computing the r-table of u'''
# init locc & r[0,..]:
for a in A: locc[a]=0; r[0,a]=0
# fill rest of r & maintain locc[..]:
for i from 1 to |u|:
    b = u[i]; locc[b] = i;
    for a in A:
        if a == b:
            r[i,a] = 1 + r[i-1,a]
        else:
            r[i,a] = min(r[i-1,a], 1 + r[locc[a],b])
\end{cminted}
\end{center}

\begin{corollary}
\label{coro-algo-h}
$h(u)$ can be computed in bilinear time
$O(|A|\cdot |u|)$.
\end{corollary}
\begin{proof}
The $r$-table and the $\ell$-table of $u$ are computed in bilinear
time as we just explained. Then one finds $\max_{a\in
A}\max_{i=0,\ldots,|u|} r(u(0,i),a)+\ell(a,u(i,|u|))+1$ by looping
over these two tables. As stated in \Cref{prop-hu-delta-u-u1au2}, this
gives $h(u)$.
\end{proof}
For $\rho(u)$, we compute the $r$- and $\ell$-vectors.
\begin{definition}[\protect{$r$-vector, $\ell$-vector, \cite[p.~73]{simon72}}]
The $r$-vector of $u=a_1\cdots a_m$ is $\tuple{r_1,\ldots,r_m}$
defined with $r_i=r(a_1\cdots a_{i-1}, a_i)$ for all $i=1,\ldots,m$.\\
The $\ell$-vector of $u$ is $\tuple{\ell_1,\ldots,\ell_m}$ defined
with $\ell_i=\ell(a_i,a_{i+1}\cdots a_m)$ for all $i=1,\ldots,m$.\\
\end{definition}
\begin{remark}
The \emph{attribute} of $u$ defined in
\cite[\textsection~3]{fleischer2018} is exactly the juxtaposition of
Simon's $r$- and $\ell$-vectors with all values shifted by 1.
\qed
\end{remark}
\begin{example}
Let us continue with $u=\mathtt{ABBACCBCCABAABC}$. Its $r$-vector and
$\ell$-vector are, respectively:
\[
\begin{array}{rc}
\text{$r$-vector:}    & \tuple{0, 0, 1, 1, 0, 1, 1, 2, 3, 1, 2, 2, 3, 3, 2}\\
\text{$\ell$-vector:} & \tuple{3, 4, 3, 2, 4, 3, 2, 2, 1, 2, 1, 1, 0, 0, 0}
\end{array}
\]
By summing the
two vectors, looking for a maximum value, and adding 1, we quickly obtain
\[
\max_{i=1,\ldots,|u|}
r\bigl(u(0,i-1),u(i)\bigr) \:+\: \ell\bigl(u(i),u(i,|u|)\bigr) \:+\:1 = 5 \:,
\]
which provides $\rho(u)$ as stated in \Cref{prop-rho-charac}.
\qed
\end{example}
One could extract the $r$- and $\ell$-vectors from the $r$- and
$\ell$-tables but there is a faster way.

The following algorithm that computes the $r$-vector of $u$ is
extracted from the algorithm in~\cite{barker2020} that computes the
canonical representative of $u$ modulo $\sim_k$. We refer to
\cite{barker2020} for its correctness. Its running time is $O(|A|+|u|)$
since there is a linear number of insertions in the stack $\texttt{L}$  and all the
positions read from $\texttt{L}$ are removed except the last read.
\begin{center}
\begin{cminted}{python}
'''Algorithm computing the r-vector of u'''
# init locc & L
for a in A: locc[a]=0;
L.push(0)
# fill rest of r & maintain locc[..]:
for i from 1 to |u|:
    a = u[i];
    while (head(L) >= locc[a]):
        j = L.pop()
    r[i] = 1+r[j] if j>0 else 0
    L.push(j)
    L.push(i)
    locc[a]=i
\end{cminted}
\end{center}
With a mirror algorithm, the $\ell$-vector is also computed in linear time.
\begin{corollary}
\label{coro-algo-rho}
$\rho(u)$ can be computed in linear time $O(|A| + |u|)$.
\end{corollary}



\section{Arch factorizations and the case of periodic words}
\label{sec-arch-and-h-rho}

In this section we analyse periodicities in the arch decomposition of
$u^n$ and deduce an algorithm for $h(u^n)$ and $\rho(u^n)$ that runs
in time $O(|A|^2|u|+\log n)$.


Let $A$ be some alphabet. An $A$-arch (or more simply an ``arch''
when $A$ is understood) is a word $s\in A^*$ such that $s$ contains
all letters of $A$ while none of its strict prefixes does. In
particular the last letter of $s$ occurs only once in $s$. A
\emph{co-arch} is the mirror image of an arch. The arch factorization
of a word $w\in A^*$ is the unique decomposition $w=s_1\cdot s_2
\cdots s_m\cdot t$ such that $s_1,\ldots,s_m$ are arches and $t$,
called the \emph{rest} of $w$, is a suffix that does not contain all
letters of $A$~\cite{hebrard91}. If its rest is empty, we say that $w$
is \emph{fully arched}.

For example, $\mathtt{ABBACCBCCABAABC}$ used in \Cref{ex-w1}
factorizes as $\mathtt{ABBAC}\cdot \mathtt{CBCCA}\cdot \mathtt{BAABC}
\cdot \epsilon$, with 3 arches and an empty rest. It is fully arched.

Reconsidering the $r$-table from \Cref{ex-w1} with the
arch factorization perspective, we notice that, at the beginning of
each arch, the	value of $r(i,a)$ coincides with the arch number:
{\includeversion{WITHARCHES}
\begin{center}
\scalebox{1.0}{
\begin{tikzpicture}[auto,x=5mm,y=3mm,anchor=mid,baseline,node distance=1.3em]
{
\def\Wc{0.92}
\def\ixh{2.7}
\def\alphah{2.9}
\def\wlh{1.5}
\def\wlet{\ta, \tb, \tb, \ta, \tc, \tc, \tb, \tc, \tc,
\ta, \tb, \ta, \ta, \tb, \tc}
\def\riA{0, 1, 1, 1, 2, 1, 1, 1, 1, 1, 2, 2, 3, 4, 4, 3}
\def\riB{0, 0, 1, 2, 2, 1, 1, 2, 2, 2, 2, 3, 3, 3, 4, 3}
\def\riC{0, 0, 0, 0, 0, 1, 2, 2, 3, 4, 2, 2, 2, 2, 2, 3}

\node[left] at (-1.0,\ixh) {\tiny $i$};
\foreach \i in {0,...,15} \node at ({\i*\Wc},\ixh) {\tiny \i};

\node[left] at (-1.0,\wlh) {$w$};
\foreach \val [count=\i] in \wlet \node at ({(\i-0.5)*\Wc},\wlh) {$\val$};

\node[left] at (-1.0,0) {$r(i,\ta)$};
\foreach \val [count=\i] in \riA \node at ({(\i-1)*\Wc},0) {\val};

\node[left] at (-1.0,-1.3) {$r(i,\tb)$};
\foreach \val [count=\i] in \riB \node at ({(\i-1)*\Wc},-1.3) {\val};

\node[left] at (-1.0,-2.6) {$r(i,\tc)$};
\foreach \val [count=\i] in \riC \node at ({(\i-1)*\Wc},-2.6) {\val};

{\tikzstyle{every path}=[draw,thick,-]
\path (-0.8,\ixh) -- (-0.8,-2.9);
\path (-3.2,0.85) -- (15.5*\Wc-0.15,0.85);
}

\begin{WITHARCHES}
\foreach \i in {0, 5, 10, 15}
\fill[rounded corners,very nearly transparent] (-0.4+\i*\Wc,0.8) rectangle (0.4+\i*\Wc,-3.4);
\end{WITHARCHES}

\begin{WITHARCHES}
{\tikzstyle{every path}=[thin,-Latex,shorten <=1mm]
\def\xtr{0.3}
\path (0*\Wc,\alphah) edge[bend left=30] node [above] {$\alpha$} (5*\Wc,\alphah);
\path (5*\Wc,\alphah) edge[bend left=30] (10*\Wc,\alphah);
\path (10*\Wc,\alphah) edge[bend left=30] (15*\Wc,\alphah);
}
\end{WITHARCHES}

}
\end{tikzpicture}

}
\end{center}}

There is in fact a more general phenomenon at work:
\begin{lemma}
\label{lem-r-arch-shift}
For any word $u\in A^*$, letter $a$ in $A$, and $A$-arch $s$:
\begin{equation}
\label{eq-r-arch-shift}
r(s \, u,a)=1+r(u,a)\:.
\end{equation}
\end{lemma}
{\renewcommand\endproof{\doendproof}
\begin{proof}
By induction on the length of $u$.
We  consider two cases.	 \\
Case 1: If $a$ does not occur in $u$ then $r(u,a)=0$ so the right hand
side of \eqref{eq-r-arch-shift} is $1$. Since $s$ is an arch it can be
factored as $s=s_1 a s_2$ with $a$ not occurring in $s_2$. Then $r(s
u,a)=1+\min_{b\in a s_2u}r(s_1,b)$. Necessarily, the last letter of
$s$, call it $c$, occurs in $a s_2u$ and since the last letter of an
arch occurs only once in the arch, $c$ does not occur in $s_1$, i.e.,
$r(s_1,c)=0$, entailing
$\min_{b\in a s_2u}r(s_1,b)=0$ and thus $r(s u,a)=1$ as needed to
establish \eqref{eq-r-arch-shift}.

\noindent
Case 2: If $a$ occurs in $u$, then
\begin{align}
\notag
1+r(u,a) &= 1+1+\min_{b\in a u_2}r(u_1,b)
\\
\shortintertext{for a factorization $u=u_1a u_2$ of $u$}
\tag*{\qed}
& \eqby{\text{i.h.}} 1+\min_{b\in a u_2} r(s u_1,b) = r(s u,a) \:.
\end{align}
\end{proof}
}
\begin{corollary}
\label{coro-r-arch-shift}
Let $v,u,t\in A^*$. If $v$ is fully arched with $k$ arches then
\begin{xalignat}{2}
r(v \, u,t) &= k + r(u,t)\:,
&
\ell(t,u \, v^\tR) &= k + \ell(t,u)\:.
\end{xalignat}
\end{corollary}

\subsection{Arch-jumping functions}

Seeing how Simon's $r$ and $\ell$ functions are connected to the
arches and co-arches of a word, the arch-jumping functions
from~\cite{SV-words2023} will be helpful.

\begin{definition}[$\alpha$ and $\beta$: arch-jumping functions]
Fix some alphabet $A$ and some word $w$ over $A$. For a position
$i\in\Cuts(w)$ we let $\alpha(i)$ be the smallest $j>i$ such that
$w(i,j)$ is an arch. Note that $\alpha(i)$ is undefined if $w(i,|w|)$
does not contain all letters of $A$.

Symmetrically, we let $\beta(i)$ be the largest $j<i$ such that $w(j,i)$
is a co-arch. This too is a partial function.
\end{definition}
The following picture shows $\alpha$ and $\beta$ on
 $w=\mathtt{ABBACCBCCABAABC}$ from \Cref{ex-w1}: $\alpha(2)=5$,
$\alpha(3)=7$, $\alpha$ is undefined
on $\{13,14,15\}$ and $\beta$ on $\{0,1,2,3,4\}$.

\begin{center}
\begin{tikzpicture}[auto,x=5mm,y=3mm,anchor=mid,baseline,node distance=1.3em]
{

{\tikzstyle{every path}=[draw,thick,-]
\draw (0,-1.0) -- (15,-1.0) -- (15,1.0) -- (0,1.0) -- cycle; 
}

\path (0,1) edge[dashed,-] (0,-1);
\path (1,1) edge[dashed,-] (1,-1);
\path (2,1) edge[dashed,-] (2,-1);
\path (3,1) edge[dashed,-] (3,-1);
\path (4,1) edge[dashed,-] (4,-1);
\path (5,1) edge[dashed,-] (5,-1);
\path (6,1) edge[dashed,-] (6,-1);
\path (7,1) edge[dashed,-] (7,-1);
\path (8,1) edge[dashed,-] (8,-1);
\path (9,1) edge[dashed,-] (9,-1);
\path (10,1) edge[dashed,-] (10,-1);
\path (11,1) edge[dashed,-] (11,-1);
\path (12,1) edge[dashed,-] (12,-1);
\path (13,1) edge[dashed,-] (13,-1);
\path (14,1) edge[dashed,-] (14,-1);
\path (15,1) edge[dashed,-] (15,-1);

\path (0,1) edge[dashed,-] (0,-2.8);
\path (5,1) edge[dashed,-] (5,-3.0);
\path (10,1) edge[dashed,-] (10,-3.0);
\path (15,1) edge[dashed,-] (15,-2.8);
\node[font=\small] at (0,-3.3) {$0$};
\node[font=\small] at (5,-3.5) {$5$};
\node[font=\small] at (10,-3.5) {$10$};
\node[font=\small] at (15,-3.3) {$15$};

{\tikzstyle{every node}=[]
\node at (0.5,0)  {$\ta$};
\node at (1.5,0)  {$\tb$};
\node at (2.5,0)  {$\tb$};
\node at (3.5,0)  {$\ta$};
\node at (4.5,0)  {$\tc$};
\node at (5.5,0)  {$\tc$};
\node at (6.5,0)  {$\tb$};
\node at (7.5,0)  {$\tc$};
\node at (8.5,0)  {$\tc$};
\node at (9.5,0)  {$\ta$};
\node at (10.5,0)  {$\tb$};
\node at (11.5,0)  {$\ta$};
\node at (12.5,0)  {$\ta$};
\node at (13.5,0)  {$\tb$};
\node at (14.5,0)  {$\tc$};
}

{\tikzstyle{every path}=[thin,-Latex,shorten <=1mm]
\path (0,1.2) edge [bend left=30] node [above] {$\alpha$} (5,1.2);
\path (1,1.2) edge [bend left=30]                         (5,1.2);
\path (2,1.2) edge [bend left=30] 		          (5,1.2);
\path (3,1.2) edge [color=blue,bend left=30] 		  (7,1.2);
\path (4,1.2) edge [color=red,bend left=30] 		  (10,1.2);
\path (5,1.2) edge [color=red,bend left=30] 		  (10,1.2);
\path (6,1.2) edge [color=red,bend left=30] 		  (10,1.2);
\path (7,1.2) edge [bend left=30] 		          (11,1.2);
\path (8,1.2) edge [bend left=30] 		          (11,1.2);
\path (9,1.2) edge [color=blue,bend left=30] 		  (15,1.2);
\path (10,1.2) edge [color=blue,bend left=30] 		  (15,1.2);
\path (11,1.2) edge [color=blue,bend left=30] 		  (15,1.2);
\path (12,1.2) edge [color=blue,bend left=30] 		  (15,1.2);

\path (15,-1.2) edge [bend left=30] node [below] {$\beta$} (12,-1.2);
\path (14,-1.2) edge [color=red,bend left=30]              (8,-1.2);
\path (13,-1.2) edge [color=red,bend left=30]              (8,-1.2);
\path (12,-1.2) edge [color=red,bend left=30]              (8,-1.2);
\path (11,-1.2) edge [color=red,bend left=30]              (8,-1.2);
\path (10,-1.2) edge [color=blue,bend left=30]             (6,-1.2);
\path (9,-1.2)  edge [bend left=30]                        (3,-1.2);
\path (8,-1.2)  edge [bend left=30]                        (3,-1.2);
\path (7,-1.2)  edge [bend left=30]                        (3,-1.2);
\path (6,-1.2)  edge [color=red,bend left=30]              (2,-1.2);
\path (5,-1.2)  edge [color=red,bend left=30]              (2,-1.2);

}

}
\end{tikzpicture}

\end{center}

\begin{lemma}[\cite{SV-words2023}]
\label{lem-alpha-beta-facts}
When the values are defined, the
following inequalities hold:
\begin{align}
i+|A|	  & \leq \alpha(i)		\:,
\\
\alpha(i) & \leq \alpha(i+1)		\:,
\\
i	  & \leq \beta\alpha(i)		\:,
\\
\alpha(i) & = \alpha\beta\alpha(i)	\:,
\\
  i \leq \beta^n\alpha^n(i)
	  & \leq \beta^{n+1}\alpha^{n+1}(i)
	    \leq \alpha(i)		\:.
\end{align}
\end{lemma}

The arch factorization $w=s_1\cdot s_2\cdots s_m\cdot t$ of $w$ can be
defined in terms of $\alpha$: $m$ is
the largest number such that $\alpha^{m}(0)$ is defined, each $s_i$ is
$w\bigl(\alpha^{i-1}(0),\alpha^i(0)\bigr)$ and $r=w(\alpha^m(0),|w|)$~\cite{SV-words2023}.
Co-arch factorizations can be defined similarly in terms of the
$\beta$ function.

\subsection{Arch factorization of periodic words}
\label{ssec-periodic-words}
\label{ssec-arch-un}

We now turn to \emph{periodic words}, of the form $u\cdot u \cdots u$,
i.e., $u^n$, where $n>0$ is the number of times $u$ is repeated. We
let $L\eqdef|u|$ denote the length of $u$. Our first goal is to
exhibit periodic patterns in the
arch factorization of $u^n$.


Assume that $u\neq\epsilon$, with $\alphabet(u)=A$. In order to study
the arch factorization of $u^n$ as a function of $n$, we set
$w=u^\omega$ and consider the (infinite) arch factorization
$u^\omega=s_1\cdot s_2\cdots s_m\cdots$: since $w$ is infinite and
since all the letters in $A$ occur in $u$, $\alpha$ is defined
everywhere over $\Nat$. For any $k\in\Nat$, we write $\lambda_k$ for
$\alpha^k(0)$, i.e., the cumulative length of $w$'s first $k$ arches.

Note that, over $u^\omega$, $\alpha(i+L)=\alpha(i)+L$ since $w$ is a
periodic word.	We say that \emph{$p\in\Nat$ is an arch-period for
$u^\omega$ starting at $i$} if there exists $k$ such that
$\alpha^k(i)\equiv \alpha^{k+p}(i) \mod{L}$. Such a period must exist
for any $i\in\Nat$: the sequence
$\alpha^0(i),\alpha^1(i),\ldots,\alpha^L(i)$ contains two values
$\alpha^{k}(i)$ and $\alpha^{k'}(i)$ that are congruent modulo $L$
and, assuming $k<k'$, one can pick $p=k'-k$.

Note that, since $\alpha(i+L)=\alpha(i)+L$ for any $i$, having
$\alpha^{k}(i)\equiv\alpha^{k+p}(i) \mod{L}$ entails
$\alpha^{k'}(i)\equiv\alpha^{k'+p}(i) \mod{L}$ for all $k'\geq k$.  In
fact, the \emph{span} $\Delta$, defined as
$\alpha^{k+p}(i)-\alpha^k(i)$, does not depend on $k$ once $k$ is
large enough.
\begin{proposition}
\label{prop-pu-exists}
There exists some integer $p_u>0$ such that, for any $i\in\Nat$, the set
of arch-periods starting at $i$ are exactly the multiples of
$p_u$.\\
Consequently $p_u$ is called \emph{the} arch-period of $u$.
\end{proposition}
\begin{ARXIV}
\begin{proof}
We first show that, for a fixed starting point $i$, the arch periods are all
multiple of a same base period.

For this, assume that $p_1<p_2$ are two arch-periods starting at $i$
and let $q=p_2-p_1$.  For $k$ large enough, one has both $\alpha^k(i)
\equiv \alpha^{k+p_2}(i) \mod{L}$ and $\alpha^{k+q}(i) \equiv
\alpha^{k+q+p_1}(i) \bigl(= \alpha^{k+p_2}(i)\bigr) \mod{L}$,
entailing $\alpha^k(i)\equiv\alpha^{k+q}(i)\mod{L}$. Thus $q$ is an
arch-period too.  Finally, the gcd of all arch-periods (starting at
$i$) is an arch-period too. Since clearly all the multiples of an
arch-period are arch-periods, our first claim is proved.

\medskip

Now proving that the arch-period does not depend on $i$ uses the
monotonicity properties listed in \Cref{lem-alpha-beta-facts}.

Pick some $i>0$ and assume that $p_i$ is the minimal period starting
from $i$ while $p_0$ is the minimal period starting from $0$, with
spans $\Delta_i$ and $\Delta_0$, respectively. Then for all $k$ large
enough, $\alpha^{k+p_0}(0) = \alpha^{k}(0)+\Delta_0$ and
$\alpha^{k+p_i}(i) = \alpha^{k}(0)+\Delta_i$. Let us pick some $k$
large enough and pick $k'$ such that
\begin{gather}
\label{eq-def-k'}
\alpha^{k'}(0)\leq\alpha^k(i)<\alpha^{k'+1}(0)
\:,
\\
\shortintertext{entailing}
\label{eq-def-k'-cont}
\alpha^k(i)<\alpha^{k'+1}(0)\leq \alpha^{k+1}(i)
\:.
\end{gather}
With \Cref{eq-def-k'}, the monotonicity of $\alpha$ entails
\begin{align*}
\alpha^{k'+p_0p_i}(0)&=\alpha^{k'}(0)+p_i\Delta_0
\leq
\alpha^{k+p_0p_i}(i)\\
&=\alpha^k(i)+p_0\Delta_i
\leq
\alpha^{k'+1+p_0p_i}(i)=\alpha^{k'+1}(0)+p_i\Delta_0
\:,
\\
\shortintertext{thus}
\alpha^{k'}(0) &\leq \alpha^k(i)+\bigl[p_0\Delta_i-p_i\Delta_0\bigr] \leq
\alpha^{k'+1}(0)\:.
\end{align*}
Since $\alpha^{k'+1}(0)-\alpha^{k'}(0)\leq L$ there are two cases:
\begin{enumerate}
\item if
$\alpha^{k'+1}(0)=\alpha^{k'}(0)+L$ then $1$ is a period and $p_0=1$.
\item
otherwise $\alpha^{k'+1}(0)<\alpha^{k'}(0)+L$ and, combining with
\Cref{eq-def-k'}, $|p_0\Delta_i-p_i\Delta_0|<L$. Since $\Delta_0$ and
$\Delta_i$ are multiple of $L$, we conclude that $p_0\Delta_i =
p_i\Delta_0$. But this requires $p_0=p_i$ and $\Delta_0 = \Delta_i$
since larger periods cannot have shorter spans.
\end{enumerate}
In both cases $p_0$ divides $p_i$. Now one proves that $p_i$ divides
$p_0$ by focusing on \Cref{eq-def-k'-cont}, from which one derives
$\alpha^k(i)\leq \alpha^{k'+1}(0)+ [p_i\Delta_0-p_0\Delta_i]\leq
\alpha^{k+1}(i)$ and proceeds as just before.

Finally, $p_0$ and $p_i$ coincide since each of them divides the other.
\end{proof}
\end{ARXIV}
\begin{example}
\label{exmp-pu-exists}
The following picture illustrates the case where $u=\mathtt{AABBCC}$
and $L=6$.  Starting at $i=0$, one has $\alpha(0)=\lambda_1=5$ and
$\alpha^4(0)=\lambda_4=17$, so $\alpha^1(0) \equiv \alpha^4(0)
\mod{L}$ and $p_u=3$ is the arch-period.
\qed
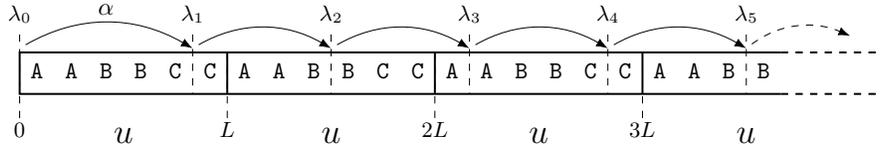
\begin{center}
\scalebox{0.92}{
\begin{tikzpicture}[auto,x=5mm,y=3mm,anchor=mid,baseline,node distance=1.3em]
{
\def\Wu{6}
\def\Lz{0}
\def\Li{\Lz+\Wu}
\def\Lii{\Li+\Wu}
\def\Liii{\Lii+\Wu}
\def\Liiii{\Liii+\Wu}
\def\Lend{\Liiii+1.1}

{\tikzstyle{every path}=[draw,thick,-]
\path (\Liiii-2,-1.0) -- (\Lz,-1.0) -- (\Lz,1.0) -- (\Liiii-2,1.0); 
\path (\Li,-1.0)    -- (\Li,1.0);
\path (\Lii,-1.0)   -- (\Lii,1.0);
\path (\Liii,-1.0)  -- (\Liii,1.0);
\path[dashed] (\Liiii-2,-1.0) -- (\Lend,-1.0);
\path[dashed] (\Liiii-2,1.0) -- (\Lend,1.0);
}
{\tikzstyle{every node}=[node font=\Large]
\node at ({(\Lz+\Li)/2},-3) {$u$};
\node at ({(\Li+\Lii)/2},-3) {$u$};
\node at ({(\Lii+\Liii)/2},-3) {$u$};
\node at ({(\Liii+\Liiii)/2},-3) {$u$};
}
\node[font=\small] at (\Lz,-2.8) {$0$};
\path (\Lz,1) edge[dashed,-] (\Lz,-2.2);
\node[font=\small] at (\Li,-2.8) {$L$};
\path (\Li,1) edge[dashed,-] (\Li,-2.2);
\node[font=\small] at (\Lii,-2.8) {$2L$};
\path (\Lii,1) edge[dashed,-] (\Lii,-2.2);
\node[font=\small] at (\Liii,-2.8) {$3L$};
\path (\Liii,1) edge[dashed,-] (\Liii,-2.2);

\def\shifta{1.8}
\def\shiftb{3.4}
\def\shiftc{0.4}
\def\shiftinu{\Wu/6}
\def\shiftai{\Wu/12}
\def\shiftaii{\shiftai+\shiftinu}
\def\shiftbi{\shiftaii+\shiftinu}
\def\shiftbii{\shiftbi+\shiftinu}
\def\shiftci{\shiftbii+\shiftinu}
\def\shiftcii{\shiftci+\shiftinu}
{\tikzstyle{every node}=[]
\node at ({\Lz+\shiftai},0)  {$\ta$};
\node at ({\Lz+\shiftaii},0) {$\ta$};
\node at ({\Lz+\shiftbi},0)  {$\tb$};
\node at ({\Lz+\shiftbii},0) {$\tb$};
\node at ({\Lz+\shiftci},0)  {$\tc$};
\node at ({\Lz+\shiftcii},0) {$\tc$};

\node at ({\Li+\shiftai},0)  {$\ta$};
\node at ({\Li+\shiftaii},0) {$\ta$};
\node at ({\Li+\shiftbi},0)  {$\tb$};
\node at ({\Li+\shiftbii},0) {$\tb$};
\node at ({\Li+\shiftci},0)  {$\tc$};
\node at ({\Li+\shiftcii},0) {$\tc$};

\node at ({\Lii+\shiftai},0)  {$\ta$};
\node at ({\Lii+\shiftaii},0) {$\ta$};
\node at ({\Lii+\shiftbi},0)  {$\tb$};
\node at ({\Lii+\shiftbii},0) {$\tb$};
\node at ({\Lii+\shiftci},0)  {$\tc$};
\node at ({\Lii+\shiftcii},0) {$\tc$};

\node at ({\Liii+\shiftai},0)  {$\ta$};
\node at ({\Liii+\shiftaii},0) {$\ta$};
\node at ({\Liii+\shiftbi},0)  {$\tb$};
\node at ({\Liii+\shiftbii},0) {$\tb$};
}
{\tikzstyle{every path}=[thin,-Latex,shorten <=1mm]
\def\xtr{0.3}
\node[font=\small] at (\Lz,2.8) {$\lambda_0$};
\path (\Lz,2.2) edge[dashed,-] (\Lz,-1);
\def\Lambdai{\Lz+\shiftci+\shiftinu/2}
\path (\Lz,1.2) edge [bend left=30] node [above] {$\alpha$} (\Lambdai,1.2);
\node[font=\small] at (\Lambdai,2.8) {$\lambda_1$};
\path (\Lambdai,2.2) edge[dashed,-] (\Lambdai,-1);
\def\Lambdaii{\Li+\shiftbi+\shiftinu/2}
\path (\Lambdai,1.2) edge [bend left=30] (\Lambdaii,1.2);
\node[font=\small] at (\Lambdaii,2.8) {$\lambda_2$};
\path (\Lambdaii,2.2) edge[dashed,-] (\Lambdaii,-1);
\def\Lambdaiii{\Lii+\shiftai+\shiftinu/2}
\path (\Lambdaii,1.2) edge [bend left=30] (\Lambdaiii,1.2);
\node[font=\small] at (\Lambdaiii,2.8) {$\lambda_3$};
\path (\Lambdaiii,2.2) edge[dashed,-] (\Lambdaiii,-1);
\def\Lambdaiv{\Lii+\shiftci+\shiftinu/2}
\path (\Lambdaiii,1.2) edge [bend left=30] (\Lambdaiv,1.2);
\node[font=\small] at (\Lambdaiv,2.8) {$\lambda_4$};
\path (\Lambdaiv,2.2) edge[dashed,-] (\Lambdaiv,-1);
\def\Lambdav{\Liii+\shiftbi+\shiftinu/2}
\path (\Lambdaiv,1.2) edge [bend left=30] (\Lambdav,1.2);
\node[font=\small] at (\Lambdav,2.8) {$\lambda_5$};
\path (\Lambdav,2.2) edge[dashed,-] (\Lambdav,-1);
\path (\Lambdav,1.2) edge [dashed,bend left=30] (\Lambdav+3,1.8);
}

}
\end{tikzpicture}

}
\end{center}
\end{example}
Recall that if $u$ can be factored as
$u=u_1u_2$ then $u_2u_1$ is a \emph{conjugate} of $u$.
\begin{proposition}
\label{prop-pv-puR}
~\\
If $v$ is a conjugate of $u$ then $u$ and $v$ have the same
arch-period.\\
If $v$ is the mirror of $u$ then $u$ and $v$ have the
same arch-period.
\end{proposition}
\begin{ARXIV}
\begin{proof}
The result is clear when $v$ is a conjugate since starting at $i=0$ in
$v^\omega$ is equivalent to starting at some other position in
$u^\omega$, so $p_u=p_v$ is just a rewording of \Cref{prop-pu-exists}.

The case where $v=u^\tR$ is more interesting.  Rather than explicitly
considering the arch-period of $u^\tR$, we'll investigate the
periodicity of the backwards function $\beta$ on an bi-infinite
extension $w=u^{\omega^*+\omega}$ of $u$.  In this setting, $\alpha$
and $\beta$ are defined everywhere on $\Rel$. Let us assume that $p$ is the
arch-period for $u^\omega$, so that $\alpha^p(i)\equiv i \mod{L}$ for
some $i$.  \Cref{lem-alpha-beta-facts} now entails
\[
i \leq \beta^p\alpha^p(i)\leq \beta^{2 p}\alpha^{2 p}(i)\leq \cdots\leq\beta^{n p}\alpha^{n p}(i)
  \leq \beta^{(n+1)p}\alpha^{(n+1)p}(i) \leq \cdots
  \leq \alpha (i) \:.
\]
There must be some $n$ large enough so that $\beta^{n p}\alpha^{n
p}(i) = \beta^{n p+p}\alpha^{n p+p}(i)$. Now, writing $\Delta$ for
the span $\alpha^p(i)-i$, one has
\begin{align*}
\beta^{n p}\alpha^{n p}(i)
&=
\beta^{n p+p}\alpha^{n p+p}(i)
=
\beta^{n p+p}\alpha^{n p}\alpha^p(i)
=
\beta^{n p+p}\alpha^{n p}(i+\Delta)
\\
&=
\beta^{n p+p}(\alpha^{n p}(i)+\Delta)
=
\beta^{n p+p}\alpha^{n p}(i)+\Delta
=
\beta^p\bigl(\beta^{n p}\alpha^{n p}(i)\bigr)+\Delta
\:.
\end{align*}
Hence $\beta^p(i')\equiv i'\mod{L}$ for $i'=\beta^{n p}\alpha^{n p}(i)$,
so $p$ is an arch-period for $u^\tR$.

The same reasoning works when swapping the roles of $u$ and $u^\tR$, so we
conclude that the arch-period for $u$ and the arch-period for $u^\tR$
divide one another. Hence they coincide.
\end{proof}
\end{ARXIV}

Note that while $p_u$ does not depend on the starting point $i$, the
smallest $k$ such that $\alpha^k(i)\equiv\alpha^{k+p_u}(i)$ usually
does. In \Cref{exmp-pu-exists} one has $k=1$ when starting from
$i=0$. But $\alpha^3(1)=13\equiv 1=\alpha^0(1)$, so $k=0$ works when
starting from $i=1$.

In the following, we shall always start from $0$: the smallest $k$
such that $\alpha^k(0)\equiv\alpha^{k+p_u}(0)\mod{L}$ is denoted by
$K_u$ and we further define $T_u=\lambda_{K_u}=\alpha^{K_u}(0)$ and
$\Delta_u=\lambda_{K_u+p_u}-\lambda_{K_u}=\alpha^{K_u+p_u}(0)-\alpha^{K_u}(0)$.
Note that $\Delta_u$ is a multiple of $L$, and we let
$\Delta_u=\delta_u L$. Together, $T_u$ and $\Delta_u$ are called the
\emph{transient} and the \emph{span} of the periodic arch
factorization.	The \emph{slope} $\sigma_u$ is $\delta_u/p_u$: after
the transient part, moving forward by $p_u$ arches in $w$ is advancing
through $\delta_u$ copies of $u$.  In the above example, we have
$p_u=3$, $T=5$ and $\Delta=2L=12$, hence $\sigma=\frac{2}{3}$ (here
and below, we omit the $u$ subscript when this does not cause
ambiguities).  \\

The reasoning proving the existence of an arch-period for $u$ shows
that at most $L$ arches have to be passed before we find
$\alpha^{k+p}(i) \equiv \alpha^k(i)\mod{L}$, so $k+p\leq L$, entailing
$K_u+p_u\leq L$ and $\delta_u\leq L$.  However, while $p_u=L$ is
always possible, this does not lead to an $L^2$ upper bound for the
span $\Delta_u$. One can show the following:
\begin{proposition}[Bounding span and transient]
\label{prop-delta-A-1}
For any $u\in A^*$, $T_u+\Delta_u\leq (|A|+1)\cdot L$.
\end{proposition}
\begin{ARXIV}
\begin{proof}
For every $m=1,\ldots,n,\ldots$, we let $k_m$ be the smallest arch
number such that $\lambda_{k_m}=\alpha^{k_m}(0)>(m-1)L$, i.e., such
that $\lambda_{k_m}$ is inside the $m$-th copy of $u$ in
$u^\omega$. The sequence $k_1<k_2<\cdots<k_n<\cdots$ is well-defined
and strictly increasing, starting with $k_1=1$.  Write
$w=s_1s_2\cdots$ for the arch-factorization of $w$. Since
$w(\lambda_{k_m})$ is the last letter of $s_{k_m}$, it is the first
occurrence of that letter in $s_{k_m}$ hence also in the $m$-th $u$
factor. (See following picture where only the first occurrences of
each letter in $u$ are depicted.  In that illustration, one has
$\lambda_{k_2} \equiv \lambda_{k_4} \mod{L}$.)
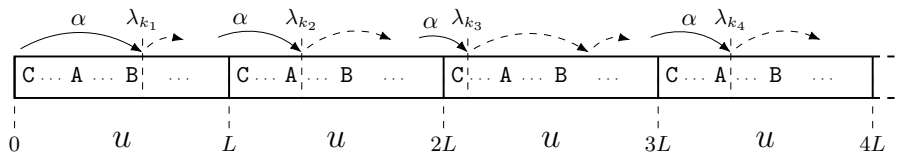
\begin{center}
\scalebox{0.92}{
\begin{tikzpicture}[auto,x=5mm,y=3mm,anchor=mid,baseline,node distance=1.3em]
{
\def\Wx{0}
\def\Wu{6.2}
\def\Lz{0}
\def\Li{\Lz+\Wx}
\def\Lii{\Li+\Wu}
\def\Liii{\Lii+\Wu}
\def\Liv{\Liii+\Wu}
\def\Lv{\Liv+\Wu}
\def\Lend{\Lv+1.1}

{\tikzstyle{every path}=[draw,thick,-]
\path (\Lv,-1.0) -- (\Lz,-1.0) -- (\Lz,1.0) -- (\Lv,1.0); 
\path (\Li,-1.0)    -- (\Li,1.0);
\path (\Lii,-1.0)   -- (\Lii,1.0);
\path (\Liii,-1.0)  -- (\Liii,1.0);
\path (\Liv,-1.0) -- (\Liv,1.0);
\path (\Lv,-1.0) -- (\Lv,1.0);
\path[dashed] (\Lv,-1.0) -- (\Lend,-1.0);
\path[dashed] (\Lv,1.0) -- (\Lend,1.0);
}
{\tikzstyle{every node}=[node font=\Large]
\node at ({(\Li+\Lii)/2},-3) {$u$};
\node at ({(\Lii+\Liii)/2},-3) {$u$};
\node at ({(\Liii+\Liv)/2},-3) {$u$};
\node at ({(\Liv+\Lv)/2},-3) {$u$};
}
{\tikzstyle{every node}=[font=\footnotesize]
\path ({\Li},0) edge[dashed,-] ({\Li},-2.8);
\path ({\Lii},0) edge[dashed,-] ({\Lii},-2.8);
\path ({\Liii},0) edge[dashed,-] ({\Liii},-2.8);
\path ({\Liv},0) edge[dashed,-] ({\Liv},-2.8);
\path ({\Lv},0) edge[dashed,-] ({\Lv},-2.8);
\node at ({\Li},-3.3) {$0$};
\node at ({\Lii},-3.3) {$L$};
\node at ({\Liii},-3.3) {$2L$};
\node at ({\Liv},-3.3) {$3L$};
\node at ({\Lv},-3.3) {$4L$};
}

\def\shifta{1.8}
\def\shiftb{3.4}
\def\shiftc{0.4}
{\tikzstyle{every node}=[]
\node at ({\Li+\shifta},0) {$\ta$};
\node at ({\Li+\shiftb},0) {$\tb$};
\node at ({\Li+\shiftc},0) {$\tc$};
\node at ({\Lii+\shifta},0) {$\ta$};
\node at ({\Lii+\shiftb},0) {$\tb$};
\node at ({\Lii+\shiftc},0) {$\tc$};
\node at ({\Liii+\shifta},0) {$\ta$};
\node at ({\Liii+\shiftb},0) {$\tb$};
\node at ({\Liii+\shiftc},0) {$\tc$};
\node at ({\Liv+\shifta},0) {$\ta$};
\node at ({\Liv+\shiftb},0) {$\tb$};
\node at ({\Liv+\shiftc},0) {$\tc$};
\node[font=\tiny] at ({\Li+(\shiftc+\shifta)/2},0) {$\cdots$};
\node[font=\tiny] at ({\Li+(\shifta+\shiftb)/2},0) {$\cdots$};
\node[font=\tiny] at ({\Li+(\shiftb+\Wu)/2},0) {$\cdots$};
\node[font=\tiny] at ({\Lii+(\shiftc+\shifta)/2},0) {$\cdots$};
\node[font=\tiny] at ({\Lii+(\shifta+\shiftb)/2},0) {$\cdots$};
\node[font=\tiny] at ({\Lii+(\shiftb+\Wu)/2},0) {$\cdots$};
\node[font=\tiny] at ({\Liii+(\shiftc+\shifta)/2},0) {$\cdots$};
\node[font=\tiny] at ({\Liii+(\shifta+\shiftb)/2},0) {$\cdots$};
\node[font=\tiny] at ({\Liii+(\shiftb+\Wu)/2},0) {$\cdots$};
\node[font=\tiny] at ({\Liv+(\shiftc+\shifta)/2},0) {$\cdots$};
\node[font=\tiny] at ({\Liv+(\shifta+\shiftb)/2},0) {$\cdots$};
\node[font=\tiny] at ({\Liv+(\shiftb+\Wu)/2},0) {$\cdots$};
}
{\tikzstyle{every path}=[thin,-Latex,shorten <=1mm]
\def\xtr{0.3}
\def\Lambdaki{\Li+\shiftb+\xtr}
\path (\Lz,1.2) edge [bend left=30] node [above] {$\alpha$} (\Lambdaki,1.2);
\node[font=\small] at (\Lambdaki,2.8) {$\lambda_{k_1}$};
\path (\Lambdaki,2.2) edge[dashed,-] (\Lambdaki,-1);
\path (\Lambdaki,1.2) edge [dashed,bend left=30] (\Lii-1.3,1.8);
\def\Lambdakii{\Lii+\shifta+\xtr}
\path (\Lii-0.5,1.8) edge [bend left=30] node [above,pos=0.3] {$\alpha$} (\Lambdakii,1.2);
\node[font=\small] at (\Lambdakii,2.8) {$\lambda_{k_2}$};
\path (\Lambdakii,2.2) edge[dashed,-] (\Lambdakii,-1);
\path (\Lambdakii,1.2) edge [dashed,bend left=30] (\Liii-1.5,1.8);
\def\Lambdakiii{\Liii+\shiftc+\xtr}
\path (\Liii-0.9,1.8) edge [bend left=30] node [above,pos=0.3] {$\alpha$} (\Lambdakiii,1.2);
\node[font=\small] at (\Lambdakiii,2.8) {$\lambda_{k_3}$};
\path (\Lambdakiii,2.2) edge[dashed,-] (\Lambdakiii,-1);
\path (\Lambdakiii,1.2) edge [dashed,bend left=30] (\Lambdakiii+3.5,1.2);
\path (\Lambdakiii+3.5,1.2) edge [dashed,bend left=30] (\Lambdakiii+4.7,1.8);
\def\Lambdakiv{\Liv+\shifta+\xtr}
\path (\Liv-0.4,1.8) edge [bend left=30] node [above] {$\alpha$} (\Lambdakiv,1.2);
\node[font=\small] at (\Lambdakiv,2.8) {$\lambda_{k_4}$};
\path (\Lambdakiv,2.2) edge[dashed,-] (\Lambdakiv,-1);
\path (\Lambdakiv,1.2) edge [dashed,bend left=30] (\Lv-1.5,1.8);
}

}
\end{tikzpicture}

}
\end{center}
Since each letter of $A$ has a single first occurrence in $u$, the
$\lambda_{k_m}$'s can only take at most $|A|$ different values modulo
$L$. Therefore by the pigeonhole principle, there are two
values $1\leq m<m'\leq |A|+1$ with $\lambda_{k_m}\equiv
\lambda_{k_{m'}} \mod{L}$ and $\lambda_{k_{m'}} - \lambda_{k_{m}}
\leq |A|L$. We deduce $p_u\leq k_{m'}-k_m\leq |A|$ and
$T_u+\Delta_u=\lambda_{K_u+p_u}\leq \lambda_{k_{m'}}\leq (|A|+1)\cdot L$.
\end{proof}
\end{ARXIV}

\subsection{Piecewise complexity of periodic words}
\label{ssec-h-rho-un}

\begin{theorem}
\label{thm-h-rho-un}
Assume $\alphabet(u)=A$ and write $L$ for $|u|$. Further let $T$,
$\Delta$ ($=\delta L$) and $p$ be the transient, span and arch-period associated with $u$, and
$T'$ be the transient associated with $u^\tR$.
\\
If $n\geq \frac{T+T'}{L}$ then
$h(u^{n+\delta})=h(u^{n})+p$ and
$\rho(u^{n+\delta})=\rho(u^{n})+p$.
\end{theorem}
\begin{ARXIV}
\begin{proof}
Write $w$ for $u^{n}$ and $w'$ for $u^{n+\delta}$ and consider the $r$
and $\ell$-tables for $w'$. For $i\geq T$ one has $r(i+\Delta,a) =
r(i,a)+p$. Symmetrically one has $\ell(a,i+\Delta) = \ell(a,i)-p$ when
$i+\Delta\leq |w'|-T'$.  This implies that any maximal
$r(w'(0,i),a)+\ell(a,w'(i,|w'|))+1$ can be realised
with $i\leq T+\Delta$ or with $i\geq |w'|-T'-\Delta$.  The same
relative positions exist in $w$, and they lead to some
$r(w(0,i),a)+\ell(a,w(i,|w|))+1$ that differ by $p$.
\end{proof}
\end{ARXIV}

\Cref{thm-h-rho-un} leads to a simple and efficient algorithm for
computing $h(u^n)$ and $\rho(u^n)$ when $n$ is large.  We first
compute $p_u$, $\Delta_u$, $T_u$ by factoring $u^{|A|}$ into arches.
We obtain $T_{u^\tR}$ in a similar way. We then find the largest $m$
such that $(n-m)\delta_u L\geq T_u+T_{u^\tR}+\Delta_u$.	 Writing
$n_0$ for $n-mp_u\delta_u$, we then compute $h(u^{n_0})$ and
$\rho(u^{n_0})$ using the algorithms from \Cref{sec-algo}. Finally
we use $h(u^n)=h(u^{n_0})+m p$ and
$\rho(u^n)=\rho(u^{n_0})+m p$.

Note that it is not necessary to compute the transients since we can
replace them with the $|A|\cdot L$ upper bound. However we need $p$ and
$\delta$, which can be obtained in time $O(|A|\cdot |u|)$ thanks to
the bound from \Cref{prop-delta-A-1}. Computing $h(u^{n_0})$ takes
time $O(|A|^2\cdot |u|)$ since $n_0$ is in $O(|A|)$, thanks again to
the bound on $T$ and $\Delta$. Finally the algorithm runs in time
$O(|A|^2|u|+\log n)$, hence in linear time when $A$ is fixed.




\section{Conclusion}
\label{sec-concl}

In this paper we focused on the piecewise complexity of individual
words, as captured by the piecewise height $h(u)$ and the somewhat
related minimality index $\rho(u)$, a new measure suggested
by~\cite{simon72} and that we introduce here.

These measures admit various characterisations, including
\Cref{prop-hu-delta-u-u1au2,prop-rho-charac} that can be leveraged
into efficient algorithms running in bilinear time
$O(|A|\cdot|u|)$ for $h(u)$ and linear time
$O(|A|+|u|)$ for $\rho(u)$. Our analysis further allows to
establish monotonicity and convexity properties for $h$ and $\rho$,
e.g., ``$\rho(u)\leq\rho(uv)\leq\rho(u)+\rho(v)$'', and to relate $h$
and $\rho$.

In a second part we focus on computing $h$ and $\rho$ on periodic
words of the form $u^n$.  We obtain an elegant solution based on
exhibiting periodicities in the arch factorization of $u^n$ and
as-yet-unnoticed connections between arch factorization and the side
distance functions, and propose an algorithm that runs in polynomial
time $O(|A|^2\cdot |u|+\log n)$, hence in linear time in contexts
where the alphabet $A$ is fixed.  This suggests that perhaps computing
$h$ and $\rho$ on compressed data can be done efficiently, a question
we intend to attack in future work.



\bibliographystyle{alpha}
\bibliography{../subwords}

\newcommand{\etalchar}[1]{$^{#1}$}
\begin{thebibliography}{GKK{\etalchar{+}}21}

\bibitem[BFH{\etalchar{+}}20]{barker2020}
L.~Barker, P.~Fleischmann, K.~Harwardt, F.~Manea, and D.~Nowotka.
\newblock Scattered factor-universality of words.
\newblock In {\em Proc.\ DLT 2020}, volume 12086 of {\em Lecture Notes in Computer Science}, pages 14--28. Springer, 2020.

\bibitem[BSS12]{bojanczyk2012b}
M.~Boja{\'{n}}czyk, L.~Segoufin, and H.~Straubing.
\newblock Piecewise testable tree languages.
\newblock {\em Logical Methods in Comp.\ Science}, 8(3), 2012.

\bibitem[CP18]{carton2018b}
O.~Carton and M.~Pouzet.
\newblock {Simon}'s theorem for scattered words.
\newblock In {\em Proc.\ DLT 2018}, volume 11088 of {\em Lecture Notes in Computer Science}, pages 182--193. Springer, 2018.

\bibitem[DGK08]{DGK-ijfcs08}
V.~Diekert, P.~Gastin, and M.~Kufleitner.
\newblock A survey on small fragments of first-order logic over finite words.
\newblock {\em Int.\ J.\ Foundations of Computer Science}, 19(3):513--548, 2008.

\bibitem[FK18]{fleischer2018}
L.~Fleischer and M.~Kufleitner.
\newblock Testing {Simon}'s congruence.
\newblock In {\em Proc.\ MFCS 2018}, volume 117 of {\em Leibniz International Proceedings in Informatics}, pages 62:1--62:13. Leibniz-Zentrum f{\"u}r Informatik, 2018.

\bibitem[GKK{\etalchar{+}}21]{gawrychowski2021}
P.~Gawrychowski, M.~Kosche, T.~Ko{\ss}, F.~Manea, and S.~Siemer.
\newblock Efficiently testing {Simon}'s congruence.
\newblock In {\em Proc.\ STACS 2021}, volume 187 of {\em Leibniz International Proceedings in Informatics}, pages 34:1--34:18. Leibniz-Zentrum f{\"u}r Informatik, 2021.

\bibitem[GS16]{goubault2016}
J.~Goubault{-}Larrecq and S.~Schmitz.
\newblock Deciding piecewise testable separability for regular tree languages.
\newblock In {\em Proc.\ ICALP 2016}, volume~55 of {\em Leibniz International Proceedings in Informatics}, pages 97:1--97:15. Leibniz-Zentrum f{\"u}r Informatik, 2016.

\bibitem[H{\'{e}}b91]{hebrard91}
J.-J. H{\'{e}}brard.
\newblock An algorithm for distinguishing efficiently bit-strings by their subsequences.
\newblock {\em Theoretical Computer Science}, 82(1):35--49, 1991.

\bibitem[HS19]{HS-ipl2019}
S.~Halfon and {\relax Ph}.~Schnoebelen.
\newblock On shuffle products, acyclic automata and piecewise-testable languages.
\newblock {\em Information Processing Letters}, 145:68--73, 2019.

\bibitem[KCM08]{kontorovich2008}
L.~Kontorovich, C.~Cortes, and M.~Mohri.
\newblock Kernel methods for learning languages.
\newblock {\em Theoretical Computer Science}, 405(3):223--236, 2008.

\bibitem[KKS15]{KKS-ipl2015}
P.~Karandikar, M.~Kufleitner, and {\relax Ph}.~Schnoebelen.
\newblock On the index of {Simon}'s congruence for piecewise testability.
\newblock {\em Information Processing Letters}, 115(4):515--519, 2015.

\bibitem[Kl{\'{\i}}11]{klima2011}
O.~Kl{\'{\i}}ma.
\newblock Piecewise testable languages via combinatorics on words.
\newblock {\em Discrete Mathematics}, 311(20):2124--2127, 2011.

\bibitem[KS16]{KS-csl2016}
P.~Karandikar and {\relax Ph}.~Schnoebelen.
\newblock The height of piecewise-testable languages with applications in logical complexity.
\newblock In {\em Proc.\ CSL 2016}, volume~62 of {\em Leibniz International Proceedings in Informatics}, pages 37:1--37:22. Leibniz-Zentrum f{\"u}r Informatik, 2016.

\bibitem[KS19]{KS-lmcs2019}
P.~Karandikar and {\relax Ph}.~Schnoebelen.
\newblock The height of piecewise-testable languages and the complexity of the logic of subwords.
\newblock {\em Logical Methods in Comp.\ Science}, 15(2), 2019.

\bibitem[Mat98]{matz98}
O.~Matz.
\newblock On piecewise testable, starfree, and recognizable picture languages.
\newblock In {\em Proc.\ FOSSACS '98}, volume 1378 of {\em Lecture Notes in Computer Science}, pages 203--210. Springer, 1998.

\bibitem[MT15]{masopust2015}
T.~Masopust and M.~Thomazo.
\newblock On the complexity of $k$-piecewise testability and the depth of automata.
\newblock In {\em Proc.\ DLT 2015}, volume 9168 of {\em Lecture Notes in Computer Science}, pages 364--376. Springer, 2015.

\bibitem[Pin86]{pin86}
J.-\'E. Pin.
\newblock {\em Varieties of Formal Languages}.
\newblock Plenum, New-York, 1986.

\bibitem[PP04]{perrin2004}
D.~Perrin and J.-{\'E}. Pin.
\newblock {\em Infinite words: Automata, Semigroups, Logic and Games}, volume 141 of {\em Pure and Applied Mathematics Series}.
\newblock Elsevier Science, 2004.

\bibitem[RHF{\etalchar{+}}13]{rogers2013}
J.~Rogers, J.~Heinz, M.~Fero, J.~Hurst, D.~Lambert, and S.~Wibel.
\newblock Cognitive and sub-regular complexity.
\newblock In {\em Proc.\ FG 2012 \& 2013}, volume 8036 of {\em Lecture Notes in Computer Science}, pages 90--108. Springer, 2013.

\bibitem[Sim72]{simon72}
I.~Simon.
\newblock {\em Hierarchies of Event with Dot-Depth One}.
\newblock PhD thesis, University of Waterloo, Waterloo, ON, Canada, 1972.

\bibitem[Sim75]{simon75}
I.~Simon.
\newblock Piecewise testable events.
\newblock In {\em Proc.\ 2nd GI Conf.\ on Automata Theory and Formal Languages}, volume~33 of {\em Lecture Notes in Computer Science}, pages 214--222. Springer, 1975.

\bibitem[Sim03]{simon2003}
I.~Simon.
\newblock Words distinguished by their subwords.
\newblock In {\em Proc.\ WORDS 2003}, 2003.

\bibitem[SS83]{sakarovitch83}
J.~Sakarovitch and I.~Simon.
\newblock Subwords.
\newblock In M.~Lothaire, editor, {\em Combinatorics on Words}, volume~17 of {\em Encyclopedia of Mathematics and Its Applications}, chapter~6, pages 105--142. Cambridge Univ.\ Press, 1983.

\bibitem[SV23]{SV-words2023}
{\relax Ph}.~Schnoebelen and J.~Veron.
\newblock On arch factorization and subword universality for words and compressed words.
\newblock In {\em Proc.\ WORDS 20123}, volume 13899 of {\em Lecture Notes in Computer Science}, pages 274--287. Springer, 2023.

\bibitem[Zet18]{zetzsche2018}
G.~Zetzsche.
\newblock Separability by piecewise testable languages and downward closures beyond subwords.
\newblock In {\em Proc.\ LICS 2018}, pages 929--938. ACM Press, 2018.

\end{thebibliography}

\end{document}